\documentclass{aastex63}
\usepackage{tensor}
\usepackage{graphicx}
\shorttitle{}
\shortauthors{}

\newcommand{\Msun}{M$_{\odot}$}

\newcommand{\Rsun}{R$_{\odot}$}
\newcommand{\M}{M$_{BD}$\ }

\newcommand{\OIa}{[O\,{\scriptsize I}]\,$\lambda$6300}
\newcommand{\OIb}{[O\,{\scriptsize I}]\,$\lambda$5577}

\newcommand{\Ha}{H$\alpha$}

\newcommand{\SIIa}{[S\,{\scriptsize II}]\,$\lambda$6731}
\newcommand{\SIIb}{[S\,{\scriptsize II}]\,$\lambda$4068}
\newcommand{\NII}{[NII]$\lambda$6583}

\newcommand{\F}{[FeII] 1.644$\mu$m\ }

\begin{document}

\title{Jets from the Upper Scorpius Variable Young Star System 2MASS J16075796-2040087 Via KECK/HIRES Spectro-astrometry}

\author[0000-0002-3741-9353]{Emma T. Whelan}
\affiliation{Maynooth University Department of Experimental Physics, National University of Ireland Maynooth, Maynooth, Co. Kildare, Ireland}
\affiliation{Astronomy and Astrophyics Section, School of Cosmic Physics, Dublin Institute for Advanced Studies, Dublin 2, Ireland}
\author[0000-0001-7250-074X]{Miriam Keppler}
\affiliation{Department of Astronomy and Steward Observatory, University of Arizona, Tucson, AZ, USA}
\author[0000-0001-8292-1943]{Neal J. Turner}
\affiliation{Jet Propulsion Laboratory, California Institute of Technology, 4800 Oak Grove Drive, Pasadena, CA 91109, USA}
\author[0000-0001-7962-1683]{Ilaria Pascucci}
\affiliation{Lunar and Planetary Laboratory, The University of Arizona,
Tucson, AZ 85721, USA}
\author[0000-0002-3131-7372]{Erika Hamden}
\affiliation{Department of Astronomy and Steward Observatory, University of Arizona, Tucson, AZ, USA}
\author[0000-0002-8636-3309]{Keri Hoadley}
\affiliation{The University of Iowa, Department of Physics $\&$ Astronomy, Van Allen Hall, Iowa City, IA 52242, USA }
\author[0000-0001-8060-1321]{Min Fang}
\affiliation{Purple Mountain Observatory, Chinese Academy of Sciences, 10 Yuanhua Road, Nanjing 210023,
China}


\begin{abstract}
2MASS J16075796-2040087 is a $\sim$ 5~Myr young star in Upper Sco with evidence for accretion bursts on a timescale of about 15~days and, uncommonly for its age, outflows traced by multi-component forbidden emission lines (FELs).  The accretion bursts may be triggered by a companion at $\sim$ 4.6~au.  We analyze HIRES spectra optimised for spectro-astrometry to better understand the origin of the several FEL velocity components and determine whether a MHD disk wind is present.  The FEL high velocity component (HVC) traces an asymmetric, bipolar jet $\sim$ 700~au long.  The jet's position angle (PA) $\sim$277$^{\circ}$ is not perpendicular to the disk.  The lower-velocity emission, classified previously as a disk wind low-velocity component (LVC), is found to have more in common with the HVC and overall it is not possible to identify a MHD disk wind component. The spectro-astrometric signal of the low velocity emission resembles those of jets and its density and ionisation fraction fall into the range of HVCs.  We suggest a scenario where the accretion bursts due to the close companion power the jets past the age where such activity ends around most stars.  The low-velocity emission here could come from a slow jet launched near the close companion and this emission would be blended with emission from the MHD wind.

{}

\end{abstract}

\section{Introduction}
Mass outflow is central to the formation of low mass stars and likely also plays a part in determining the conditions for planet formation \citep{Ray2021, Whelan2023a}.  Outflows in the forms of jets and winds are detected for most of the pre-main-sequence phase of stars and thus can persist for millions of years \citep{Whelan2014}.  After the first observations of outflows in star-forming regions, much focus was on large-scale protostellar jets and molecular outflows \citep{reipurthbally2001}.  The importance of small-scale jets driven by more-evolved young stars and magneto-hydrodynamic (MHD) winds was quickly realised however, due to their likely role in the removal of angular momentum from the star-disk system \citep{Frank2014, Pascucci2022}.  Classical T~Tauri stars (CTTSs) are class~II, low-mass, optically-visible pre-main sequence stars.  They have cleared most of their natal envelope and therefore offer opportunities to study jets and MHD winds down to the scales where they are launched \citep{Ray2007} with the only limit being the available angular resolution \citep{Murphy2021}.  CTTSs are typically associated with small-scale jets ($<$1~arcmin, $<$10,000~au) \citep{Dougados2000, Flores2023} but parsec-scale jets are observed in some cases \citep{McGroarty2004, Comeron2011}. 

More-evolved TTSs or class~III low-mass stars often lack jets, suggesting that at some stage during a $\sim$10-Myr window the jet switches off.  For example, \cite{nisini2018} carried out a spectroscopic survey of 131 class~II sources across three star-forming regions with ages of 1-3~Myr and found that 30$\%$ of the sample had a jet.  \cite{Fang2023} conducted a similar study of 115 class~II stars in the Upper Scorpius association (age $>$5~Myr) and found only 5~sources with a jet.  These studies offer strong evidence of the frequency and size of jets reducing with age.  

Jets from CTTSs can be identified by the detection of the so-called forbidden emission line  (FEL) high velocity component (HVC, V$>$100~kms$^{-1}$) while the FEL low velocity component (LVC, $<$30~kms$^{-1}$) is suggested to trace a disk wind \citep{Whelan2004, Fang2023}.  The \OIa\ line in particular has been investigated and the HVC is normally spatially resolved so that the morphology of the jet can be studied \citep{Simon2016, Pascucci2020, Murphy2021}.  Whether the wind responsible for the compact LVC is a MHD or photoevaporative wind is strongly debated and the main motivation of \cite{Fang2023} and other similar studies was to understand the origin of the FEL LVC. In the MHD scenario, the innermost portion of the MHD wind is collimated into a jet which is then detected as a HVC \citep{Pascucci2022}.  High-resolution spectroscopic studies revealed that the LVC could be further decomposed into a narrow and a broad component (NC, BC) \citep{Simon2016} and suggested that the BC was tracing an MHD wind but that the origin of the NC was less clear \citep{Fang2018, Banzatti2019}. 

Spectroscopic studies have also used the FEL ratios to determine conditions in the gas at the two velocities and try to pin down their origins \citep{Fang2018, Banzatti2019, Weber2020}.  Indeed the earliest studies of FEL LVCs suggested separate origins for the HVC and LVC due to the higher density in the LVC \citep{HartiganApJ1995, Kwan1988}.  More recently, \cite{Nisini2023} investigated both the kinematic and local properties of the regions traced by FELs in 36~CTTSs using high-spectral-resolution data (R$\sim$115,000).  They confirmed a higher density in the LVC and an ionisation fraction $<$0.1.  Importantly they discussed the limitations of the diagnostics and the resulting difficulties with estimating mass-loss rates in the LVCs.  \cite{Giannini2019} found the physical parameters in their smaller sample of 6~CTTS varied smoothly with velocity, evidence in favor of a common MHD origin for the HVC and LVC.

\begin{figure*}
\centering
\includegraphics[width=18cm, trim= 0cm 0cm 0cm 0cm, clip=true]{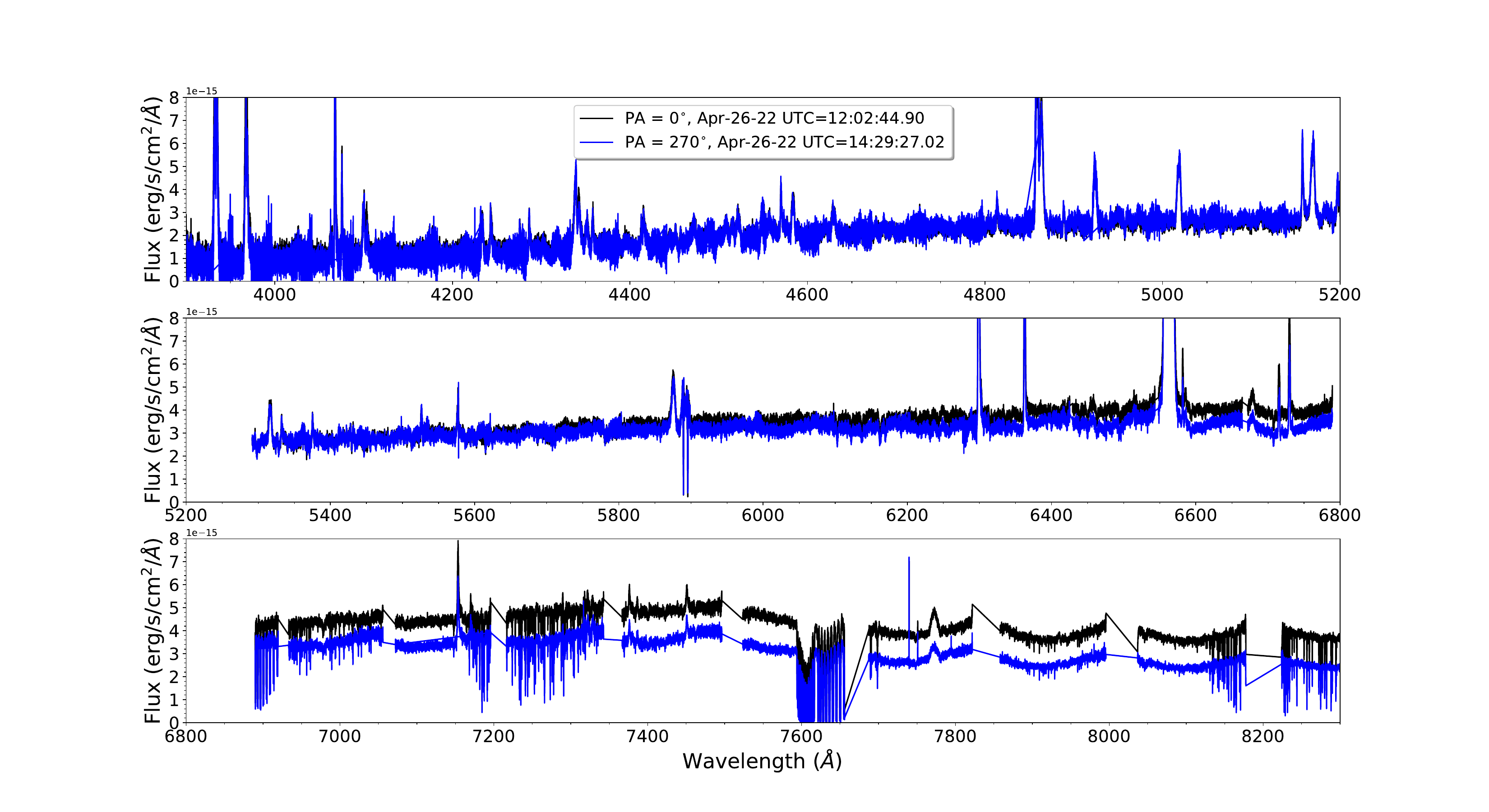}
   \caption{Full HIRES spectra of 2M160756. The spectra taken at 0$^{\circ}$ (black) and 270$^{\circ}$ (blue) are compared as they are the two spectra with the largest separation in time ($\sim$ 1~hr). Variability is seen in the continuum emission at the longer wavelengths. Between 3800~\AA\ and 5200~\AA\ there is no change in the continuum emission but between 7000~\AA\ and 8000~\AA\ the flux of the continuum emission changes by $\sim$ 1$\times$10$^{-15}$ erg/s/cm$^{2}$/\AA. \cite{Espaillat2021} discuss how the UV variability can lag behind the optical variability in young stars exhibiting variable accretion.} 
  \label{spectrum}     
\end{figure*}

\begin{deluxetable*}{llll}
\tablecaption{Average fluxes of the detected lines measured for the 90$^{\circ}$ and 270$^{\circ}$ spectra. The fluxes were measured as part of the spectral decomposition procedure and the uncertainty on the flux measurements is 10$\%$ on average. Values are reported for the accretion tracers and the outflow components. The outflow components identified in this study are the jet with a blue-shifted velocity of -70 to -80~kms$^{-1}$, the LVE with a blue-shifted velocity of -10 to -30~kms$^{-1}$ and the red-shifted jet with a velocity of $\sim$ 100~kms$^{-1}$. The profiles of the accretion tracers are presented in Figure \ref{Acc_lines} and the contribution to the H$\alpha$ line from the jet is assumed to be negligible.}
\label{fluxes}
\tablehead{
\colhead{Identification} & & &  \\
\hline
\colhead{$\lambda_{air}$}(\AA) &\colhead{Ion}  &Component  &\colhead{Flux} ($\times$ 10$^{-14}$ erg/s/cm$^{-2}$)
}
\startdata
3933.7     &Ca II   &Accretion  &3.6  \\
3968.5     &Ca II   &Accretion  &2.3  \\
4068.6     &[S II]  &Jet        &0.5  \\ 
4068.6     &[S II]  &LVE        &1.0  \\
4076.3     &[S II]  &Jet        &0.4  \\
4101.7     &H I     &Accretion  &0.9  \\
4244.0     &[Fe II] &Jet        &0.4  \\ 
4340.5     &H I     &Accretion  &0.2  \\ 
4352.8     &[Fe II] &Jet        &0.4  \\
4359.3     &[Fe II] &Jet        &0.3  \\
4814.5     &[Fe II] &Jet        &0.3  \\ 
4861.3     &H I     &Accretion  &5.4  \\
5577.3     &[O I]   &Jet        &0.1  \\ 
5577.3     &[O I]   &LVE        &0.1  \\ 
5875.9     &He I    &Accretion  &1.3  \\ 
6300.3     &[O I]   &Jet        &2.6 \\
6300.3     &[O I]   &LVE        &1.8 \\
6363.7     &[O I]   &Jet        &0.9  \\
6363.7     &[O I]   &LVE        &0.4  \\
6562.8     &H I     &Accretion  &44.2  \\
6583.5     &[N II]  &Jet        &0.1 \\
6583.5     &[N II]  &LVE        &0.1   \\
6583.5     &[N II]  &Red Jet    &0.1   \\
6716.4     &[S II]  &Jet        &0.2  \\
6716.4     &[S II]  &LVE        &0.2  \\
6716.4     &[S II]  &Red Jet    &0.1  \\
6730.8     &[S II]  &Jet        &0.3    \\ 
6730.8     &[S II]  &LVE        &0.4 \\ 
6730.8     &[S II]  &Red Jet    &0.2 \\
7155.2     &[Fe II] &Jet        &0.4  \\
7155.2     &[Fe II] &LVE        &0.1  \\
7291.5     &[Ca II] &Jet        &0.1  \\
7155.2     &[Fe II] &LVE        &0.1  \\
7377.8     &[Ni II] &Jet        &0.2 \\ 
7377.8     &[Ni II] &Red Jet    &0.1 \\ 
7388.2     &[Fe II] &Jet        &0.1\\
7452.5     &[Fe II] &Jet        &0.2  \\
\enddata

\end{deluxetable*}

What is missing from these types of high-spectral-resolution investigations is spatial information on the LVC.  The unknown geometry and height of the wind-tracing emission region is also important for constraining the mass-loss rate \citep{Fang2018, Nisini2023}.  Knowing the wind mass-loss rate can help to distinguish between a magnetic and photoevaporative origin.  Furthermore, if the efficiency of the wind is known i.e.\ the ratio of the mass outflow to accretion rates, this can be used to test the importance of the wind to the removal of angular momentum \citep{Fang2018}.  \cite{Whelan2021} have demonstrated that spatial information can be provided through spectro-astrometric study of the FEL regions.  Not only can spectro-astrometry recover spatial information on the LVC \citep{Whelan2021} but it can also be used to resolve very small-scale jets that are not directly spatially resolved \citep{Takami2001, Fedriani2018, Whelan2023b}. 

Here we present an analysis of the FELs of the CTTS 2MASS J16075796-2040087 (hereafter 2M16075796) in Upper Scorpius \citep{Luhman2020}. \cite{Fang2023} identified the \OIa\ emission line from 2M16075796 as having a HVC plus a LVC.  The motivation here is to further understand what spectro-astrometry and line diagnostics can reveal about the LVC.  2M16075796 is chosen as an interesting object due to the peculiar nature of its FEL profiles, in that the line peak-to-continuum ratio of the LVC is less than that of the HVC, while the opposite is normally the case \citep{Banzatti2019}.  A further goal of this work is to test the usefulness of KECK/HIRES spectroscopy as a complement to the proposed Astrophysics Medium Explorer Mission, Hyperion \citep{2022Hamden}.  Hyperion would determine young stars' mass outflow rates more-reliably by combining ground-based FEL spectro-astrometry, like that we demonstrate here for 2M16075796, with outflow masses determined by measuring the far-ultraviolet fluorescent line emission from the molecular hydrogen (over wavelengths 138.5–161.5~nm) that is a major constituent of the flows.

The paper is laid out as follows.  In Section~2 we discuss the target in detail along with the reduction of the data.  In Section~3 we describe the technique of spectro-astrometry.  The measured kinematics, spatial properties, and conditions in the outflows traced by the FELs are presented in Section~4 where the mass outflow and accretion rates are also estimated.  Our findings for this interesting object are summarized in Section~5 in the context of our overall understanding of FEL LVCs.

\begin{figure*}
\centering
\includegraphics[width=14cm, trim= 0cm 0cm 0cm 0cm, clip=true]{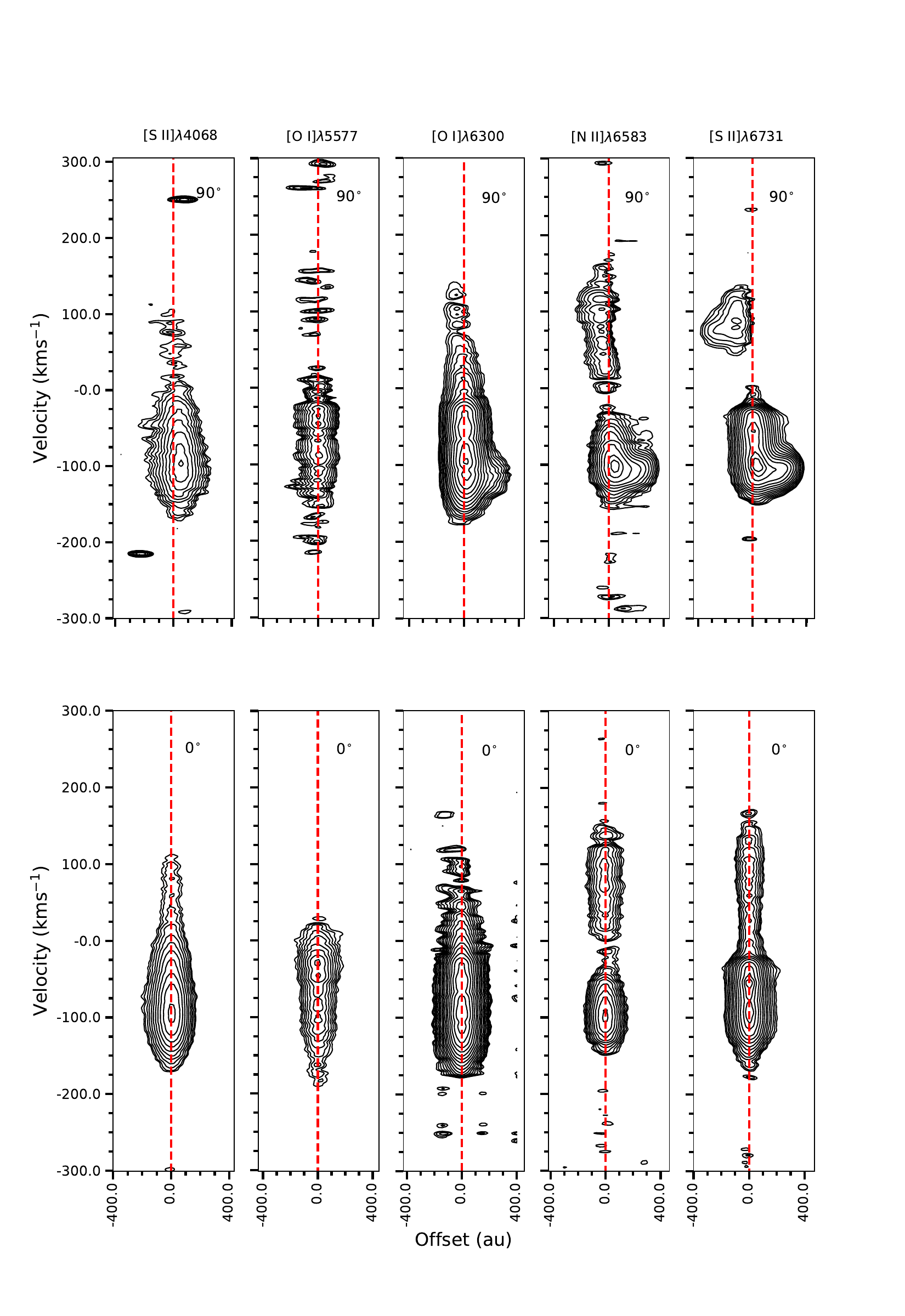}
   \caption{Continuum-subtracted position-velocity diagrams of key FELs.  Contours begin at 3~times the RMS noise and increase with an interval of $\sqrt{2}$.  The top row is from the spectrum at slit PA 90$^{\circ}$ and the bottom row from the spectrum at slit PA 0$^{\circ}$.  Both the blue- and red-shifted high-velocity emission at 90$^{\circ}$ are extended, while at 0$^{\circ}$ they are compact by eye.  Thus the jet traced by the HVC has PA close to 90$^{\circ}$.}
  \label{PVs}     
\end{figure*}

\begin{figure}
\centering
\includegraphics[width=10cm, trim= 0cm 0cm 0cm 0cm, clip=true]{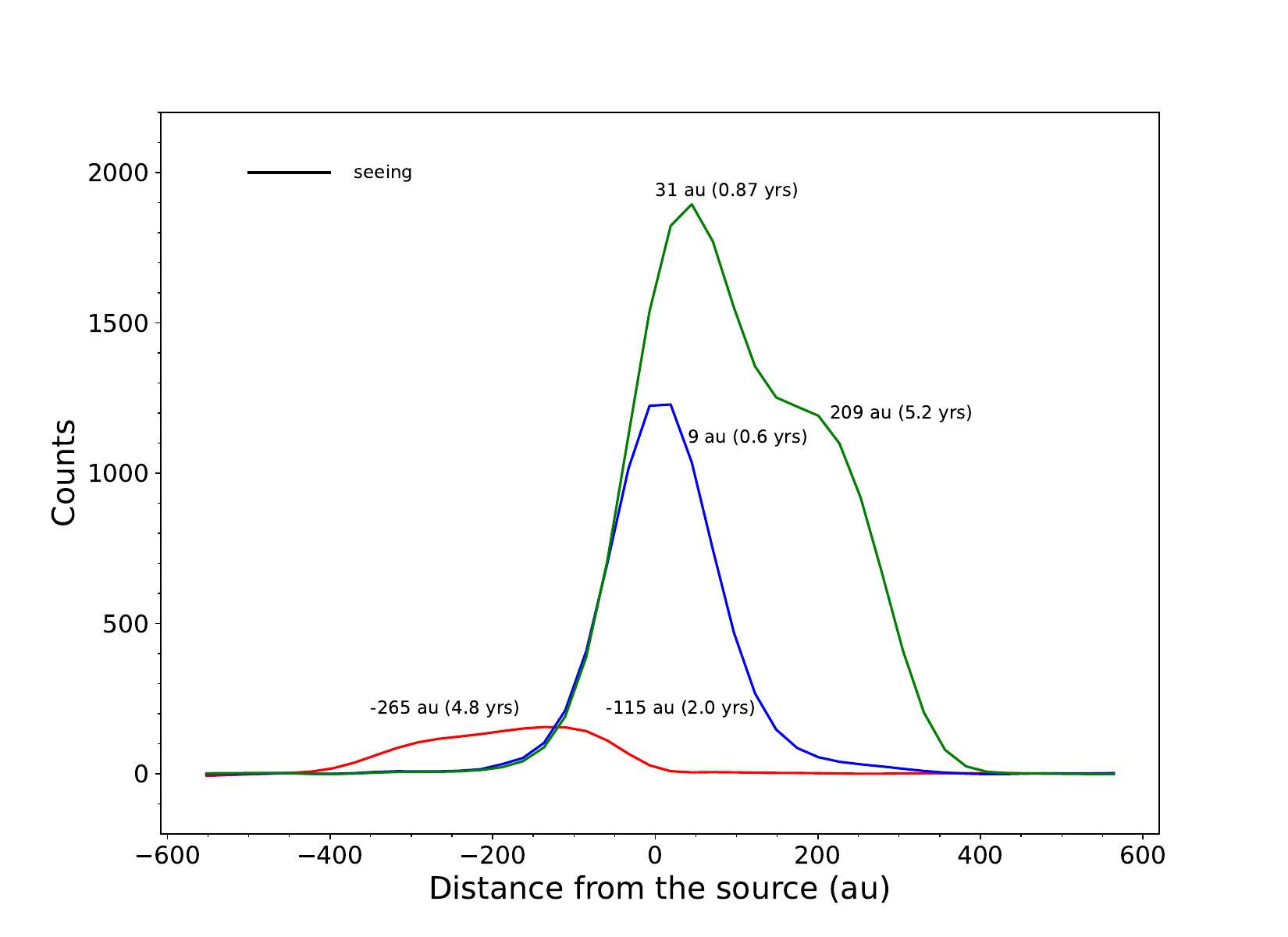}
   \caption{Outflow peaks in the \SIIa\ emission region from the spectrum with slit PA 90$^{\circ}$.  Each curve is a cut through the \SIIa\ PV diagram at the velocity of either the blue-shifted jet (green curve), the LVE (blue), or the red-shifted jet (red).  The positions of the emission peaks were combined with the tangential velocities of Table~3 to estimate the emission features' dynamical ages.  The outer knots in the red and blue jets have similar ages.  As the inner part of the red-shifted flow is hidden by the disk, it is not possible to check for a counterpart to the blue-shifted knot at 35~au.}
  \label{knots}     
\end{figure}

\begin{deluxetable*}{lccc}
\tablecaption{Velocity components identified through spectral decomposition {\bf for the FELs shown in Figures 2 and 4 and analysed with spectro-astrometry}.  From the measured radial velocities, corresponding tangential velocities are given assuming the outflow is inclined 22$^{\circ}$ with respect to the plane of the sky.  The numbers in parentheses are from fitting the LVE or jet emission alone after subtracting the other components as described in section~4.1. The uncertainty on the radial velocity measurements is 1~kms$^{-1}$.}
\label{decomp}
\tablehead{
\colhead{Emission Line} & \colhead{V$_{rad}$} (km~s$^{-1}$)   & \colhead{V$_{tan}$} (km~s$^{-1}$) &FWHM (km~s$^{-1}$)   
}
\startdata
\SIIb   &-71 (-71) &-178  &78 (78)   \\
        &-13 (-12) &-33  &130 (130)    \\
\OIb    &-70  (-71) &-175  &68 (72)   \\
        &-11 (-11) &-28  &43 (46)    \\
\OIa    &-75 (-76) &-188  &68 (64)   \\
        &-15 (-17) &-38  &41 (52)    \\
        &21  &53  &192    \\
\NII    &-80 (-80) &-200  &44 (43)  \\
        &-47 (-47) &-118  &47 (47)  \\
        &118 (118)  &295 &110 (110)  \\
\SIIa   &-76 (75) &-190  &52 (47)   \\
        &-27 (-27) &-68  &42 (42)    \\
        &101 (112) &280  &208 (57)   \\       
\enddata

\tablecomments{} 
\end{deluxetable*}
\label{velocities}

\begin{figure*}
\centering
\includegraphics[width=20cm]{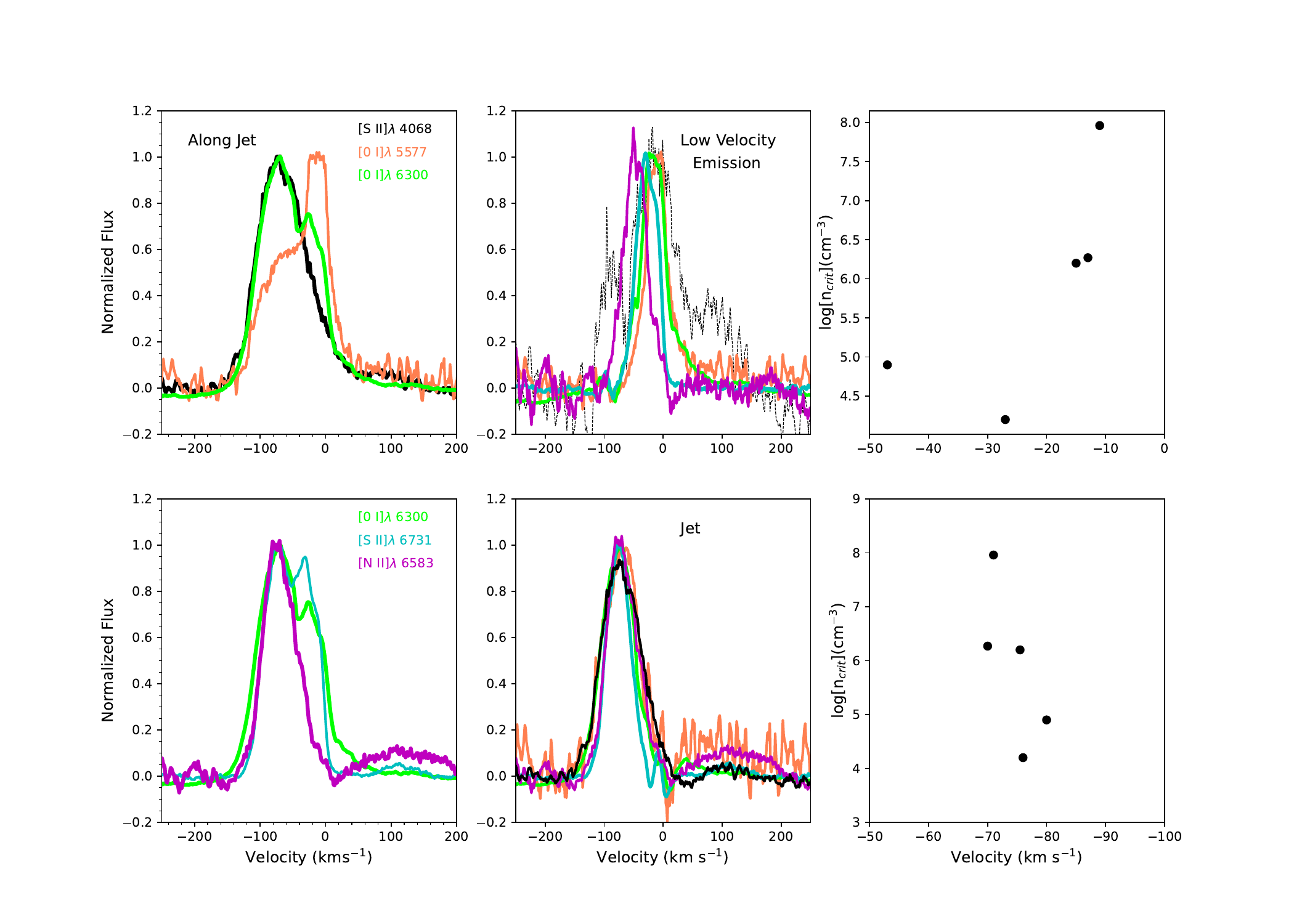}
   \caption{Comparing the FEL profiles of the lines from Figure~2.  Left column: Line profiles extracted from the source position, well away from the knots at $\sim$1\farcs5 or $\sim$200~au.  The four lines' critical densities decrease from \OIb\ to \SIIa\ while their temperatures are similar.  Two velocity components are present.  Low-velocity emission in the \OIb\ line is centered near $-15$~kms$^{-1}$ and jet emission near $-80$~kms$^{-1}$.  Middle column: LVE (upper) and jet (lower) emission in the \SIIb, \OIb, \OIa, \NII, and \SIIa\ lines after subtracting the fitted kinematic components.  The colour scheme matches the left column.  Right column: The same two components' centroid velocities versus critical density.  The LVE velocity increases with decreasing critical density, while the jet velocity changes little with critical density.}
  \label{Lines_comp}     
\end{figure*}

\begin{figure*}
\centering
\includegraphics[width=16cm, trim= 0cm 0cm 0cm 0cm, clip=true]{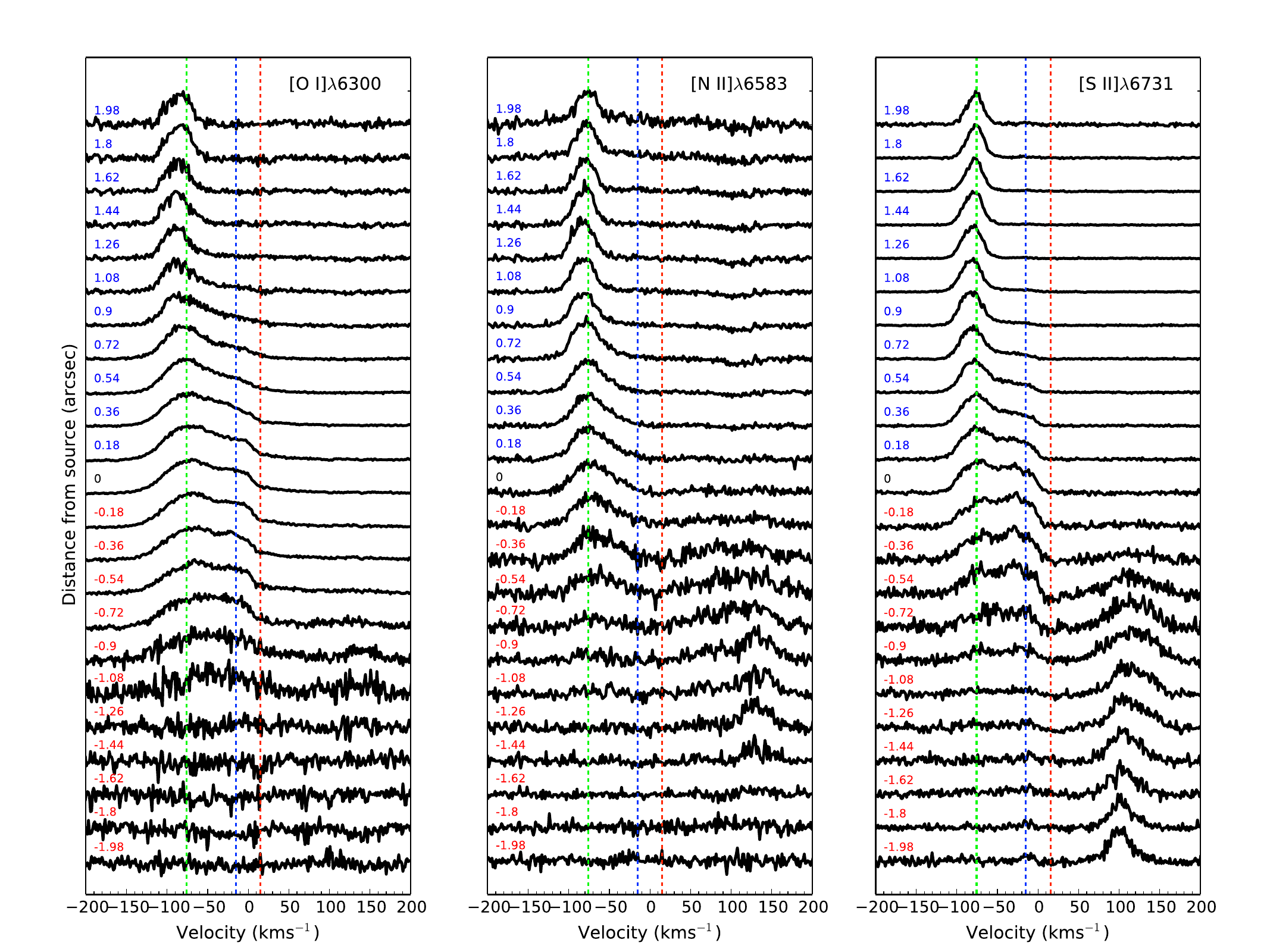}
   \caption{Key emission line profiles' variations with distance from the source.  \OIa\ and \SIIa\ show a clear LVC within 1\arcsec\ (145~au) of the source.  The blue- and red-shifted jets dominate outside 1\arcsec.  Dashed lines mark the centroid velocities for the three components measured in the line decomposition of Section~4.1.  }
  \label{waterfall}     
\end{figure*}

\begin{deluxetable*}{ccc}
\tablecaption{Dynamical ages of the jet knots and LVE.}
\label{ages}
\tablehead{
\colhead{Component} &\colhead{Distance (au)} & \colhead{Age (yrs)} 
}
\startdata
Jet blue &34.8  &0.87\\
Jet blue &208.8 &5.2 \\
Jet red  &114.6 &2.0 \\
Jet red  &265.4 &4.8 \\
LVE      &8.7   &0.6 \\
\enddata
\end{deluxetable*}

\section{Target, Observations and Analysis}\label{obs}
2M16075796 is a K4 young star in Upper Scorpius (d = 145~pc) with a strong infrared excess and it is classified as having a full accretion disk \citep{Luhman2020, Fang2023}. It is assumed here that it has an age of $\sim$ 5~Myr in line with the estimated age of Upper Scorpius. \cite{Ratzenbock2023a} report it to be a member of the $\nu$~Sco sub-region of Upper Scorpius and \cite{Ratzenbock2023b} estimate the age of this sub-region to be 4~Myr to 6~Myr.  Alternatively, \citet{Miret-Roig2022} use traceback ages to estimate an age of $\sim$ 1~Myr for $\nu$~Sco. However, they note that the perturbations of stellar orbits due to stellar feedback and dynamical interactions could account for
the reason for the $\sim$ 4~Myr difference between dynamical traceback age and the age of $\nu$~Sco estimated using other methods. To further investigate the age of our target we consider its position on a HR diagram. \cite{Cazzoletti2019} plot the Upper Scorpius sample from \cite{Barenfield2016} including 
2M16075796 on a HR diagram with 
the evolutionary tracks for different stellar masses and the relative isochrones from
Baraffe et al. (2015). 2M16075796 with Log(T$_{eff}$) = 3.57 and Log(L$_{\star}$) = -0.82 lies between the 5~Myr and 10~Myr isochrones in support of our assumption that it is an older TTS.

\cite{Barenfield2016} presented ALMA 0.88-mm continuum and CO J=3-2 observations of the disk yielding a diameter of $<$40~au in the continuum.  \cite{Barenfeld2017} fitted models to these same observations, finding a disk dust radius R$_{dust}$ $\sim$11~au, a disk inclination of 47$^{\circ}$, and aprojected major axis position angle (PA) of 0$^{\circ}$, as well as corresponding values for the CO emission (R$_{\rm CO}$ $\sim$34~au, inclination = 52$^{\circ}$, and major axis PA = 1$^{\circ}$).  \cite{Hendler2020} approached the modeling of the same datasets differently, estimating the inclination and PA at 68$^{\circ}$ and 209$^{\circ}$ respectively.  Interestingly, \cite{Cody2017} classified 2M16075796 as a burster since it exhibited an ``accretion burst''-type lightcurve in their K2 time series photometry.  The outbursts occurred on a $\sim$15-day timescale consistent with episodic accretion.

2M16075796 was first identified as a strong accretor in the low-resolution spectra published by \cite{Kraus2009}.  These spectra also revealed numerous multi-component FELs tracing outflow activity.  \cite{Fang2023} reported that the star had a mass-accretion rate log($\dot{M}_{acc}$) = -8.96.  They used spectral fitting to decompose the \OIa\ line into a HVC (V = -76.6~km s$^{-1}$, FWHM = 61~km s$^{-1}$), a LVC NC (V = -14.8~km s$^{-1}$, FWHM = 44.6~km s$^{-1}$), and a LVC BC (V = 15.1~km s$^{-1}$, FWHM = 202.5~km s$^{-1}$), where V is the peak radial velocity of the component. 

2M16075796 appears to have at least two companions.  \cite{Kraus2009} associated it with a sub-stellar companion orbiting at $\sim$3120~au, spectroscopically confirming the candidate binary identification from the 2MASS survey of \cite{Kraus2007a}.  The pair consists of a 0.7~\Msun\ star orbited by a 0.072~\Msun\ M6 dwarf at PA 20$^{\circ}$.  The H$\alpha$ equivalent width (EW) of the primary was $-357$~\AA\ and the sub-stellar companion $-14$~\AA.  A second candidate companion at PA 357.5$^{\circ}$ $\pm$ 2.7$^{\circ}$ and projected separation 31.9$\pm$3.7~mas or $\sim$4.6~au (assuming d = 145~pc) was detected by \cite{Barenfeld2019} using non-redundant aperture masking with the NIRC2 AO imager on the Keck II telescope in May 2015.  The companion's K-band magnitude was 9.95$\pm$0.25, compared to 7.81$\pm$0.02 for the primary.

\cite{Barenfeld2019} discuss the circumbinary nature of the 2M16075796 circumstellar disk with the main evidence being the companion's location within the dust disk.  One theory that connects the circumbinary disk, strong accretion and outflow activity, and the bursting behaviour is that material from the circumbinary disk is moving across the inner gap and onto both stars.  In this case it would be expected that the accretion would be modulated with a period of order of that of the binary orbit. Although the timescale of the optical variability is measured to be $\sim$15 days, much shorter than the orbital period of a binary with a projected separation of 4.6~au, \cite{Barenfeld2019} discuss how accretion onto the system being fed and maintained by streams would solve this discrepancy.


The data presented here were obtained at the Keck telescope on 2022 April 26th, using the HIRES spectrograph red arm with a slit 0.86$^{\prime\prime}$ wide and 7$^{\prime\prime}$ long and achieving a spectral resolution of R = 50000. Following the requirements for spectro-astrometry (Section~2), spectra were obtained at four slit PAs: 0$^{\circ}$, 90$^{\circ}$, 180$^{\circ}$ and 270$^{\circ}$, with exposure times of 45~mins for the first 3 PAs and 30~mins for the 270$^{\circ}$ PA. Two standard stars (HR~1675 and 2MASS~J16112057-1820549) were also observed for the purpose of flux calibration and telluric and photospheric correction. The spectra were reduced using the Mauna Kea Echelle Extraction (MAKEE) pipeline written by Tom Barlow\footnote{\url{https://sites.astro.caltech.edu/~tb/makee}}. 1D and 2D spectra were extracted for the purpose of spectral decomposition to identify the different outflow components present and spectro-astrometry to recover spatial information on the components. The 1D spectra were flux calibrated and corrected for telluric and photospheric features before the spectral decomposition was performed. The spectrum of the standard star HR~6175 was used to derive the response curve for the flux calibration. The flux calibration procedure which followed common practices as outlined in \cite{Fang2023} was applied to the spectrum of the other standard, 2MASS J16112057-1820549 and the results compared with Gaia spectra in order to test the robustness of the procedure. Very good agreement was found and the procedure was then applied to the target spectrum. The telluric correction was done using the standard HR~1675 by following the basic technique for correction of telluric features described in the HIRES Data Reduction Manual\footnote{\url{https://www2.keck.hawaii.edu/inst/common/makeewww/Atmosphere/index.html}}. The photospheric correction was done following the procedure first outlined in \cite{Hartigan1989} where a photospheric standard with spectral type similar to 2M1607579 (2MASS J16112057-1820549) was rotationally broadened, and shifted in velocity to match the
photospheric lines of the target star. Veiling is also considered and corrected line profiles were obtained by subtracting the best-fit photospheric spectra from the target
spectra. The spectral decomposition of the line profiles was then done by finding the minimum number of Gaussians needed to fit line profiles. Gaussian components were added until the best fit was achieved and the velocity centroid and FWHM of the Gaussians along with the integrated flux were returned. The uncertainty in the velocity 
centroid was calculated from 
\begin{equation}
\sigma = \frac{0.4*{\rm FWHM}}{\rm SNR},
\end{equation}
\citep{Porter2004} and the average uncertainty for the lines that were subsequently analysed with spectro-astrometry was $\sim$ 1~kms$^{-1}$. A heliocentric velocity correction was applied to the spectra in keeping with \cite{Fang2023}. For the figures in this paper all distances are given in au. A distance of 145~pc is used to convert from arcsecond to au.

\begin{figure*}
\centering
\includegraphics[width=22cm, trim= 5cm 0cm 0cm 0cm, clip=true]{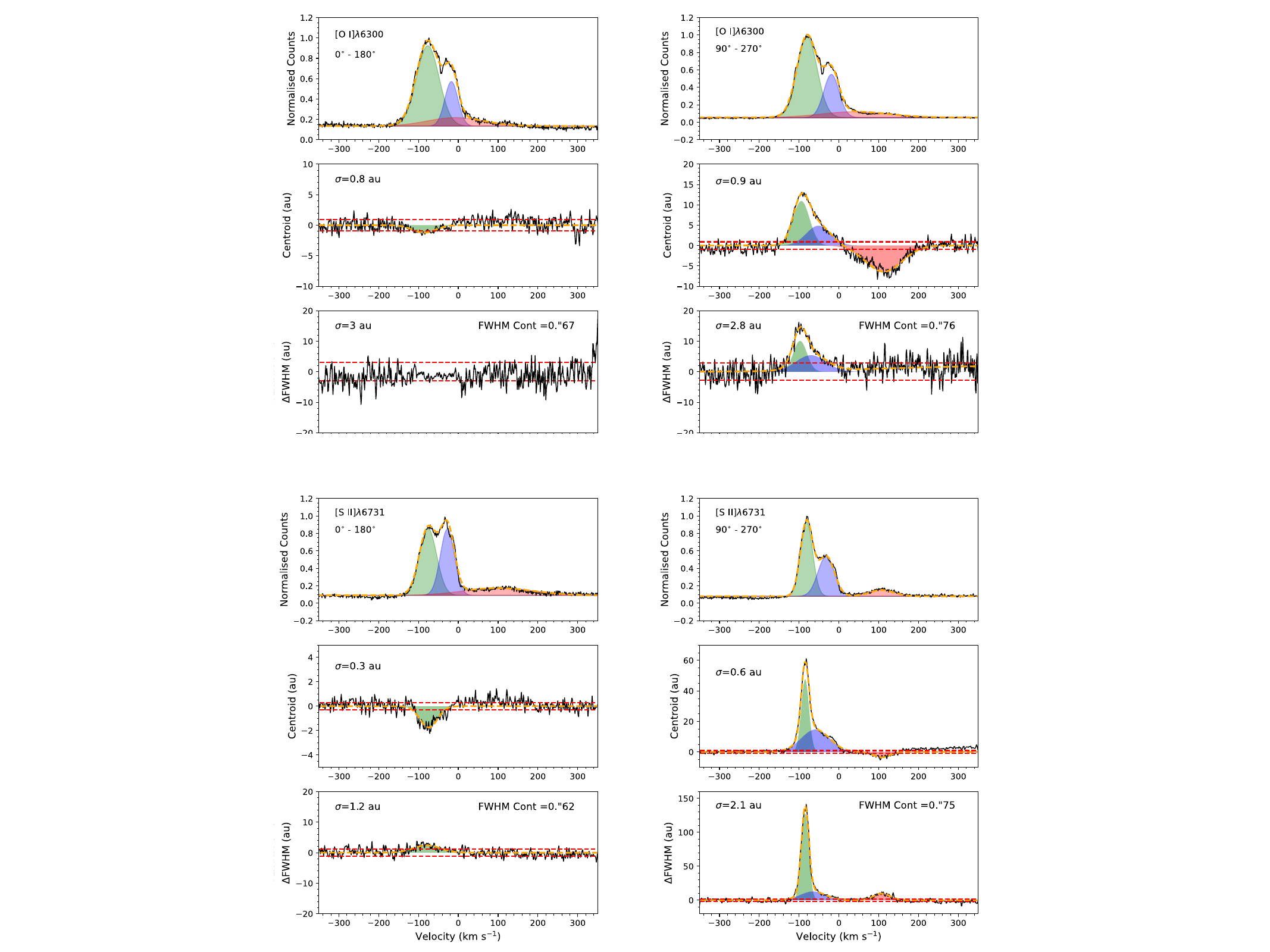}
   \caption{Spectro-astrometric analysis of the \OIa\ and \SIIa\ lines.  Results for the N-S and E-W slit PAs are presented in the left and right columns, respectively.  The \OIa\ line is in the upper two quadrants and the \SIIa\ line in the lower two.  The three panels in each quadrant show from top to bottom the line emission normalized to its peak, the centroid position, and the FWHM.  The centroids have not been corrected for continuum contamination.  Fitted Gaussian components are overplotted.  In both lines a shift in the centroid at the line peak to $\sim$ 2~au is measured in the N-S spectrum meaning that the PA of the jet is not perfectly E-W. }
  \label{SA1}     
\end{figure*}

\section{Spectro-astrometry}\label{SA}
Spectro-astrometry is a technique for extracting spatial information from spatially-unresolved emission-line regions \citep{Whelan2008}.  It has been used to great effect to study outflows, binary stars, and black holes, for example \citep{Bailey1998,Whelan2005,Gnerucci2010}.  For a 2D spectrum made up of continuum emission and an emission line region, a spatial profile is extracted at each pixel along the dispersion direction, or over a specified wavelength bin.  The centroid and full-width-half-maximum (FWHM) of each spatial profile are plotted as a function of velocity (measured with respect to the emission line rest wavelength) and in this way the spatial properties of the line region and continuum are mapped with velocity and compared. The fitting is normally done using Gaussians \citep{Whelan2021}.  The resulting plot of centroid versus velocity is referred to as a {\it position spectrum} and offers insight into the line-emitting region's displacement with respect to the continuum source and into how this displacement changes with velocity.  The plot of FWHM versus velocity constrains the size of the line-emitting region.  Commonly the FWHM measurements are plotted as $\Delta$FWHM, the FWHM of the individual spatial profile minus the average FWHM of the continuum emission.  In ground-based observations, the continuum emission's average FWHM is generally equal to the seeing \citep{Porter2004}. The power of this technique is that as the uncertainty in the centroid measurement is dependent on the signal-to-noise ratio (SNR) of the observation according to Equation 1 and therefore sub-au precision can easily be reached in the nearest star-forming regions \citep{Ponto2011}.  The uncertainty on the FWHM line measurements can be taken as the standard deviation in the continuum FWHM measurements. 

Important points to note are firstly, as the technique measures the spatial centroid of the emission line region it gives a minimum estimate of any displacement in the line region.  Secondly, if there is more than one component to the emission region under investigation, for example line and continuum emission or two blended line components, then the centroid position measured is some combination of the signals from both components.  For this reason the position spectrum of a line region is often measured before and after the subtraction of the continuum emission.  Additionally, the contaminating effect of the continuum emission can be estimated by considering the ratio of the line to continuum emission \citep{Takami2001, Whelan2021}.  Thirdly, artifacts due to tracking problems can also be an issue \citep{Brannigan2006}.  The technique of using anti-parallel slits adopted by \cite{Brannigan2006} removes these artifacts.  Specifically, the centroid measurements done at the anti-parallel slit positions are subtracted to remove the artifacts.  It is also beneficial to obtain spectra at perpendicular slit PAs as this allows one to produce a 2D spatial map of the emission region.  A typical sequence for spectro-astrometric observations thus involves 4~spectra for example at slit PAs 0$^{\circ}$, 90$^{\circ}$, 180$^{\circ}$ and 270$^{\circ}$.  For targets with a known jet or disk, that feature's orientation is often taken as the zero-point from which the four spectra are offset in 90$^\circ$ increments \citep{Whelan2021}. 

\section{Results}\label{sect:results}
In Figure \ref{spectrum} we compare the full HIRES spectra of 2M16075796 with slit PAs 0$^{\circ}$ (black) and 270$^{\circ}$ (blue).  These are the first and last observations in the sequence of four spectra.  The continuum emission decreases between the two observations but the change is most pronounced at longer wavelengths and is only seen beyond $\sim$5600~\AA.  At 7000~\AA\ the decrease in flux corresponds to a change in brightness of 0.3~mag.  The change in the continuum emission is greatest between the two slit PAs shown here but it is seen across all four spectra.  Thus the change is most likely due to the intrinsic variability of the source rather than the change in slit orientation.  Young stars' variability often differs between UV and optical wavelengths \citep{Espaillat2021,Robinson2022}.  For example, a lag between the optical and UV variations in GM~Aur may be due to a density gradient in the accretion hotspots on the star's surface \cite{Espaillat2021}. 

The HIRES spectra of 2M16075796 show numerous emission lines tracing both outflow and accretion activity, most being FELs (Table~\ref{fluxes}). As each FEL is produced under a different range of conditions, comparing the FEL regions provides information on the outflow components present \citep{Hirth1997, Simon2016}.  With this in mind, continuum-subtracted position-velocity (PV) diagrams in key FELs and at the slit PAs of 0$^{\circ}$ and 90$^{\circ}$ are shown in Figure~\ref{PVs}. 
The critical densities of the lines shown are approximately 10$^{8}$, 10$^{6}$, 10$^{5}$, and 10$^{4}$~cm$^{-3}$ for the \OIb, \OIa, \NII, and \SIIa\ lines respectively \citep{Giannini2019}.  The \SIIb\ line has a similar critical density to the \OIa\ line and a similar temperature can be assumed for all the lines \citep{Nisini2023}.  PV diagrams of [Fe~II], [Ca~II], and [Ni~II] are shown in the Appendix.  A clear HVC (hereafter referred to as the jet) is seen in all the emission line regions of Figures \ref{PVs} and \ref{PVs_B} and at both PAs.  Also generally seen is lower-velocity emission, hereafter referred to as the LVE.

Examining the PV diagrams in Figure \ref{PVs} for the 90$^{\circ}$ slit PA, the \OIa, [N~II]$\lambda$6583, and \SIIa\ high velocity emission line regions are extended to $\sim$400~au on the blue-shifted side and $\sim$350~au on the red-shifted side while none of the emission line regions at the 0$^{\circ}$ PA are extended by eye.  There is a hint of the red-shifted jet in [S~II]$\lambda$4068 but the blue-shifted emission looks compact.  The [O~I]$\lambda$5577 emission also looks compact by eye.  This suggests a jet aligned more nearly E-W than N-S and a disk inclined enough to make some of the red-shifted flow visible.  Figure \ref{knots} is a cut through the \SIIa\ PV plot at the peak velocities of the jets and the LVE.  Fitting these profiles provides an estimate of the position of the knots in the jet and the spatial peak of the LVE flow.  Combining these positions with the velocities of Table~\ref{velocities} gives the flows' dynamical ages (Table~\ref{ages}).  The outer knots in the red and blue jets have similar ages suggesting the ejections occurred about the same time.  The inner part of the red-shifted flow is obscured by the disk, making a comparison of the inner knots impossible.  However, the ages of the inner and outer knots in the blue-shifted jet suggest an interval of $\sim$4~yr between ejections.

The LVE is more difficult to decipher and in Figure \ref{Lines_comp} first column, the line profiles of the 90$^{\circ}$ slit PA FELs (normalised to the line peak) are compared to offer a clearer picture.  The spectra have been extracted from the source position.  The LVE appears as an extended wing in the [S~II]$\lambda$4068 line but as a clear primary peak sitting on top of a broader jet component in the [O~I]$\lambda$5577 line.  In the \OIa\ and \SIIa\ lines, the LVE appears as a secondary peak blended with the jet.  This secondary peak was identified as a LVC by \cite{Fang2023} for the \OIa\ line.  As will be discussed further in Section~5, past studies of the \OIa\ emission for large samples of young stars show that it is far more common for the LVC to have a higher line peak-to-continuum ratio than the HVC and therefore to appear as a primary peak in the multi-component line profiles.  Further interesting features here are the difference between the shapes of the low-velocity emission in the \OIa\ and \OIb\ lines and the fact that the jet is detected in \OIb.  Jets have been found to appear in \OIb\ far less often than in \OIa\ \citep{Fang2018}.  Finally, the \NII\ emission, which traces high excitation and therefore does not typically have a LVC, here shows a hint of the LVE in its red wing. 

In Figure~\ref{waterfall} the \OIa, [N~II]$\lambda$6583, and [S~II]$\lambda$6731 emission line profiles are plotted at several distances from the star to investigate the variation more closely.  In \OIa\ and [S~II]$\lambda$6731, the blue-shifted LVE is detected out to $\sim$ 70~au as a clear second peak.  This demonstrates that the two components are spatially separated.  Also note that in all lines, blue-shifted emission is detected in the direction of the red-shifted flow out to $\sim$ 130~au.  This could be caused by scattering of the blue-shifted jet emission along the direction of the red-shifted flow but it could also be suggestive of a binary.  In the binary scenario, the spatial centroid of the continuum emission (distance of 0 in Figure \ref{waterfall}) is not the same as the position of the star driving the jet.  However, this would point to a binary separation of $\sim$1~\arcsec\ which has not been detected for 2M16075796 and makes the scattered emission explanation more plausible.

\subsection{Kinematic Analysis}
The results of the Gaussian decomposition of the emission lines profiles of Figures \ref{PVs} and \ref{Lines_comp} are given in Table \ref{decomp} and the fits are shown in Figure 6 and in the appendix. The components identified for \OIa\ are in agreement with the results of \cite{Fang2023}. Considering all the lines the blue-shifted jet close to the source position has a radial velocity between 70~kms$^{-1}$ and 80~kms$^{-1}$. For a disk with an inclination of 68$^{\circ}$ this translates to a range in tangential velocity of 175~kms$^{-1}$ to 200~kms$^{-1}$. It is assumed here that the disk and jet are perpendicular. The red-shifted jet is clearly detected in the \SIIa\ and \NII\ lines with a tangential velocity of 280~kms$^{-1}$ to 290~kms$^{-1}$. This kind of velocity asymmetry between the two lobes is frequently seen in jets from young stars \citep{Murphy2021}. The LVE is only seen on the blue-shifted side and it makes sense that any red-shifted emission originating close to the star would be hidden by the disk. It is argued that the red-shifted component identified at V $\sim$ 21~kms$^{-1}$ in \OIa\ by the Gaussian fitting is from the red-shifted jet which is only partially detected in this line. 

 To investigate the different components further the Gaussian fits were subtracted to first leave only the LVE (Figure \ref{Lines_comp} middle column top) and then the jet emission (Figure~\ref{Lines_comp} middle column bottom). The LVE was fit again after the removal of the jet emission and these values are given in parentheses in Table~\ref{decomp}. Strong residuals of the jet still remain the \SIIb\ line after the subtraction of the LVE. In Figure \ref{Lines_comp} right column top the peak velocity is plotted as a function of critical density for the 5 lines. A clear increase in peak velocity with decreasing critical density is seen. This change in peak velocity with density has been  interpreted as a density dependence in the inner parts of an accelerating flow \citep{Giannini2019, HartiganApJ1995}. Applying the criterion that separates a HVC from a LVC (V $>$ 30~kms$^{-1}$) to the radial velocities in Table \ref{velocities} means that \SIIb, \OIb, \OIa, and \SIIa\ LVE would be classified as a LVC while the \NII\ emission is a HVC. If the same criterion is applied to the tangential velocities only the \OIb\ LVE is a clear LVC. A change in peak velocity between the different lines is not seen in the jet component before or after the subtraction of the LVE (Figure \ref{Lines_comp} bottom right). This is a sign that the jet is dominated by shocks and that there is no density dependence.  A difference in the jet component is that the \NII\ and \SIIa\ lines are approximately 33\% narrower than the \SIIb, \OIb, and \OIa\ jet components.

\subsection{Spectro-astrometric Analysis}

The next step towards understanding the 2M16075796 FELs is to analyse the emission using spectro-astrometry.  The 1-D spectro-astrometry for the FELs discussed above is presented in Figure~\ref{SA1} and in the Appendix (Figures 12, 13).  The H$\alpha$ line is also included to check for a binary signal \citep{Bailey1998}.  The centroids are measured by Gaussian fitting to the spatial profiles and the results for the anti-parallel slit positions are subtracted to remove artifacts.  The centroid measurements were done both before and after continuum subtraction and the position spectra before continuum subtraction are presented here.  The 1-$\sigma$ uncertainty is an average across the velocity range plotted.  The precision improves at the line peak due to the increase in the signal at the line.  For lines where extended emission is seen, such as \SIIa, the centroid of the inner emission is plotted.  Also shown are the emission line profiles and changes in the FWHM of spatial profiles.  The 1-D results for the \OIa\ and \SIIa\ lines are shown in Figure \ref{SA1}.  The top panel in each of the four quadrants of Figure \ref{SA1} shows the emission line profile with the kinematic components reported in Table~2 overlaid.  Green is used for the blue-shifted jet, blue for the LVE and red for the red-shifted jet emission.  The bottom panel in each quadrant shows the FWHM corresponding to the line centroid. 

\begin{figure*}
\centering
\includegraphics[width=18cm, trim= 0cm 9cm 0cm 8cm, clip=true]{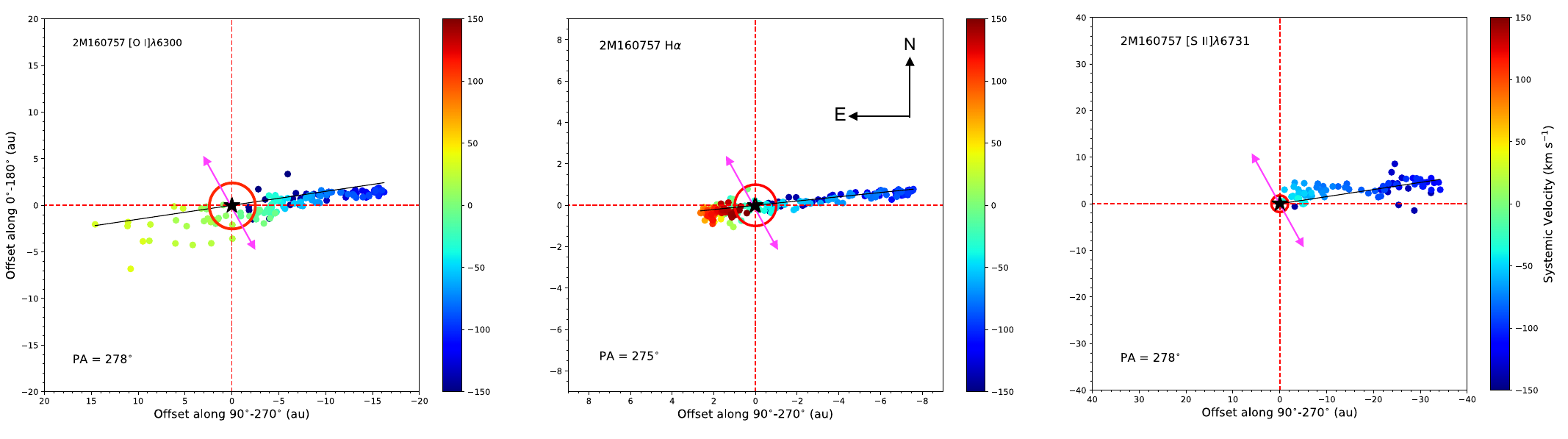}
   \caption{2-D spectro-astrometry of the \OIa, H$\alpha$, and \SIIa\ lines.  The line centroids are corrected for the continuum emission.  Magenta arrows mark the PA of the accretion disk and the red oval is the 3-$\sigma$ per-point positional uncertainty.  The PA of the blue-shifted jet is 275$^{\circ}$ in the H$\alpha$ line and 278$^{\circ}$ in the two FELs.  The red-shifted \SIIa\ jet emission is located at $>$100~au and so lies off the \SIIa\ figure.}
  \label{SA_2D}     
\end{figure*}

A centroid shift is detected in all the emission lines except the \OIb\ and at both PAs. In most cases this is accompanied by a corresponding change in the FWHM of the spatial profiles. The signal from the \OIa, \SIIa, \Ha, and \NII\ lines at 90$^{\circ}$ is primarily tracing the bipolar jet. The much smaller signal at 0$^{\circ}$ indicates that the jet PA is not exactly 90$^{\circ}$. The \OIb\ emission does not show any displacement before or after continuum subtraction and therefore is compact compared to the other lines. This is in agreement with previous spectro-astrometric studies of \OIb\ \citep{Whelan2021}. The continuum emission in the region of the \SIIb\ was significantly fainter than at the other lines and therefore it was not possible to produce a high precision position spectrum. However, results for this line are in agreement with the other lines where a displacement is measured.

To try to distinguish a difference between the signals from the jet and the LVE the same kinematic fitting that was applied to the line profiles is applied to the centroid and FWHM measurements of the \OIa\, \SIIa\ and \NII\ emission regions and over-plotted. Two components are detected in all cases which are broadly in agreement with the fitting to the line profiles. The analysis of the H$\alpha$ emission region shows only a jet signature and no evidence of a companion. This could be expected given the small separation estimated for the closer companion, if the companion is less massive and is suggestive of a substellar mass companion. The wide separation companion reported by \cite{Kraus2009} would have fallen outside the slit at all the PAs. 

\begin{figure*}
\centering
\includegraphics[width=21cm, angle=-90, trim= 0cm 0cm 0cm 0cm, clip=true]{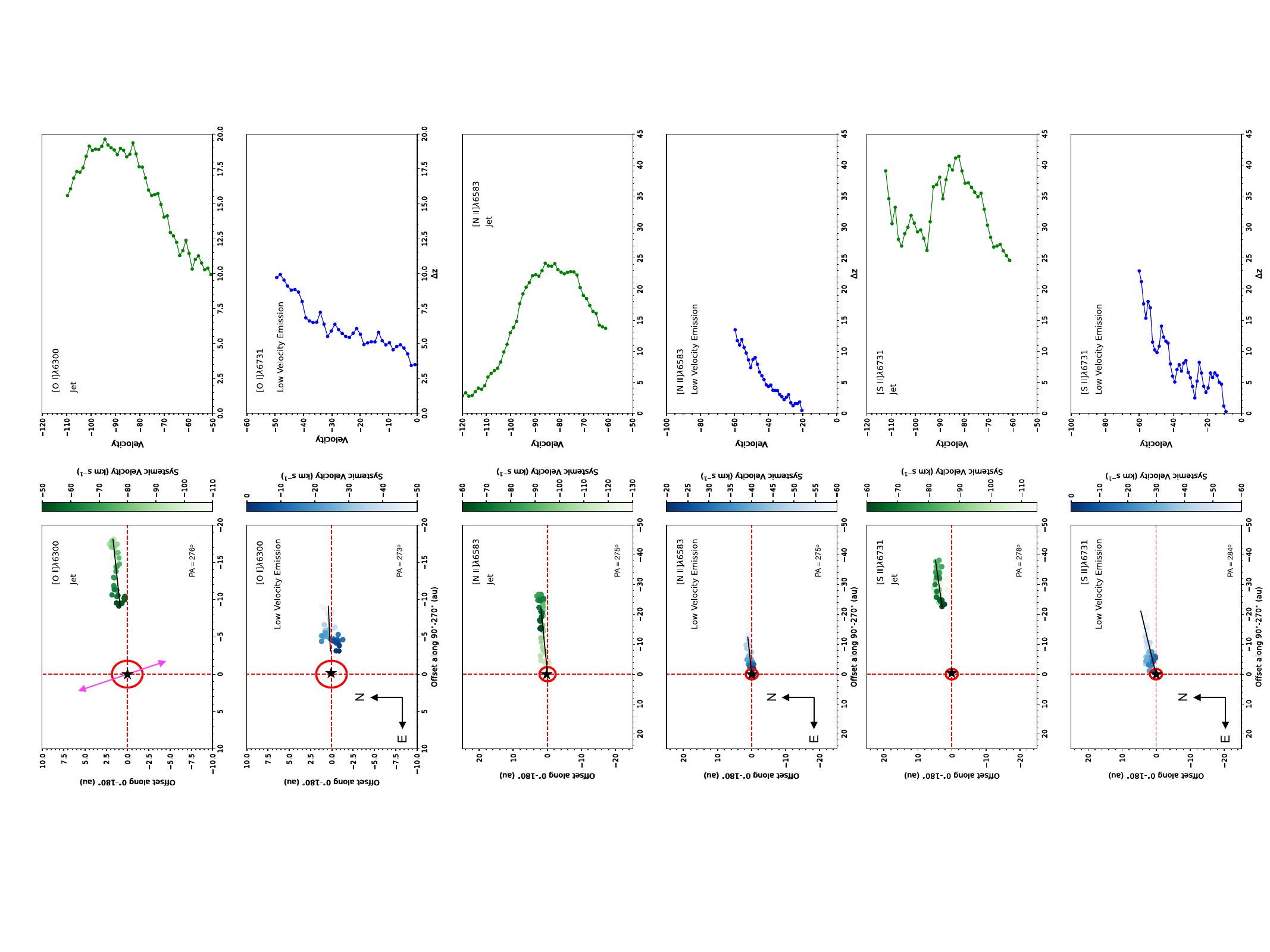}
   \caption{2D spectro-astrometry of the jet and LVE components in the \OIa, \NII\ and \SIIb\ emission (left column) and corresponding runs of outflow velocity with height above the disk (right column).  The PA of the accretion disk is marked by the magenta arrow in the \OIa\ jet panel. The red oval is the 3-$\sigma$ uncertainty.  In the velocity versus height plots the centroids are corrected for the system's inclination.  The velocities generally increase with height in the parts nearest the disk, consistent with magneto-centrifugally launched outflows.}
  \label{SA_2D_all}     
\end{figure*}

In Figures~\ref{SA_2D} and \ref{SA_2D_all}, the continuum-subtracted centroid measurements at $0^{\circ} - 180^{\circ}$ and $90^{\circ} - 270^{\circ}$ are combined to make a 2D map of the spatial properties of the emission line regions.  In each panel the PA of the accretion disk is marked by the magenta arrow and the red oval is the 3-$\sigma$ precision of the individual centroid measurements.  In Figure~\ref{SA_2D} the measurements for the full extent of the \OIa\, H$\alpha$, and \SIIa\ emission constrain the jet PA at 275$^{\circ}$ to 278$^{\circ}$, which is not perpendicular to any available measurement of the disk PA.  This could be caused by bending in the jet perhaps due to the presence of a companion.  The red-shifted \SIIa\ emission is not plotted here as it lies at $>100$~au and so is not included in the fitting once the continuum is removed.  In Figure \ref{SA_2D_all} the measurements are separated into two components based on the fitting to the line profiles, position spectra, and FWHM measurements.  Also shown in the right columns are the corresponding plots of the emission's height above the disk ($\Delta z$) with radial velocity.  The displacement along the fitted jet PA in Figure~\ref{SA_2D} is de-projected for the inclination of the system to derive $\Delta z$.  No difference in the spatial properties of the emission with velocity is identified for the two components in either line.  In both cases the results are consistent with magneto-centrifugal launching since $\Delta z$ initially increases with velocity. 


\subsection{Line Diagnostics and Outflow Efficiency}

In Table~\ref{diagnostics} we list values for key line ratios and compare against the ranges \cite{Nisini2023} found for their sample of HVCs and LVCs.  The \OIb/\OIa\  ratio probes the electron density $n_e$ and has been used to distinguish between the HVC and LVC.  For example, \cite{Banzatti2019} find values in the range $-0.8$ to $-0.23$ for the LVCs and $-1.22$ to $-0.64$ for the HVCs in their sample.  \cite{Nisini2023} report the HVCs in their sample to fall between $-1.7$ and $-1.25$ and the narrow LVCs between $-1.15$ and zero.  In 2M16075796, both the jet and the LVE have ratios consistent with the HVCs of these studies.  The [S~II]$\lambda$6716/[S~II]$\lambda$6731 ratio also samples $n_e$.  Both ratios point to a higher density in the LVE and red-shifted jet than in the blue-shifted jet.  The [S~II]$\lambda$6716/[S~II]$\lambda$6731 ratio gives $n_e\sim 3000$~cm$^{-3}$ in the blue-shifted jet and $n_e\sim 10^4$~cm$^{-3}$ in the LVE and red-shifted jet, taking a temperature of 10$^{4}$~K. 

The \NII/\OIa\ ratio can be used to constrain the outflow's degree of ionisation.  The values we measure in 2M16075796 are outside the ranges estimated for LVCs and greater than those estimated for jets.  Following the models of \cite{Nisini2023} the ratios imply ionization fractions $x_e$ between 0.5 and 1 for both the blue jet and the LVE (depending on the assumed temperature and density), with the LVE the lower of the two.  This is generally higher than observed for jets and far higher than predicted for LVCs ($x_e<0.1$) \citep{Nisini2023, Giannini2019}.  The weaker ionisation in the LVE is consistent with findings that the ionisation degree increases with the jet velocity \citep{Giannini2019}.

Predicted \SIIb/\OIa\ ratios under both a photoevaporative wind model and an MHD wind model are an order of magnitude greater than observed \citep{Nisini2023, Weber2020}.  It may be that the \OIa\ flux includes contributions from the photo-dissociation of OH and from fluorescence excited by stellar UV photons \citep{Rigliaco2013, Gangi2023}.  Another possibility is that sulfur could be depleted in the disk with respect to the solar abundance \citep{Nisini2023}.  The values in Table~\ref{diagnostics} for both the blue-shifted jet and LVE components of 2M16075796 fall into the range of values presented in \cite{Nisini2023}.

\begin{deluxetable*}{cccccc}
\tablecaption{Line ratios for the jet and LVE compared with \cite{Nisini2023} and \cite{Banzatti2019}. }
\label{diagnostics}
\tablehead{
\colhead{Ratio} & \colhead{Blue Jet} & \colhead{Low Velocity Emission}  & \colhead{Red Jet} &HVC range &LVC range
}
\startdata
$\log($[O~I]$5577/$[O~I]$6300)$    &$-1.4 $&$-1.26$&---   &$-1.25$ to $-1.7$ &$0$ to $-1.15$   \\
$\log($[S~II]$6716/$[S~II]$6731)$  &$-0.18$&$-0.3 $&$-0.3$&&  \\
$\log($[N~II]$6583/$[O~I]$6300)$   &$-1.4 $&$-1.26$&$-1  $&$-0.175$ to $-1.15$ &$<0.1$   \\
$\log($[S~II]$4068/$[O~I]$6300)$   &$-0.7 $&$-0.26$&---   &&$-0.2$ to $-0.8$ \\
 \enddata
\end{deluxetable*}

\begin{deluxetable*}{lcccc}
\tablecaption{Mass outflow rates measured from the \SIIa\ line luminosity}
\label{Mout}
\tablehead{
\colhead{Emission Line} & \colhead{$V_{\rm jet}$ (km~s$^{-1}$)}   & \colhead{$l_{\rm jet}$ (arcsec)} &$\log(\dot M_{\rm out}/$\Msun yr$^{-1})$ &$\dot M_{\rm out}$/$\dot M_{\rm acc}$ (\%) 
}
\startdata
\SIIa    &$-190$  &$2.2$  &$-8.75$  &$6.0$  \\
\SIIa    &$-68 $  &$1.0$  &$-9.13$  &$2.5$  \\
\SIIa    &$+280$  &$1.7$  &$-9.12$  &$2.6$  \\
\enddata
\end{deluxetable*}

\begin{figure*}
\centering
\includegraphics[width=14cm, trim= 0cm 0cm 0cm 0cm, clip=true]{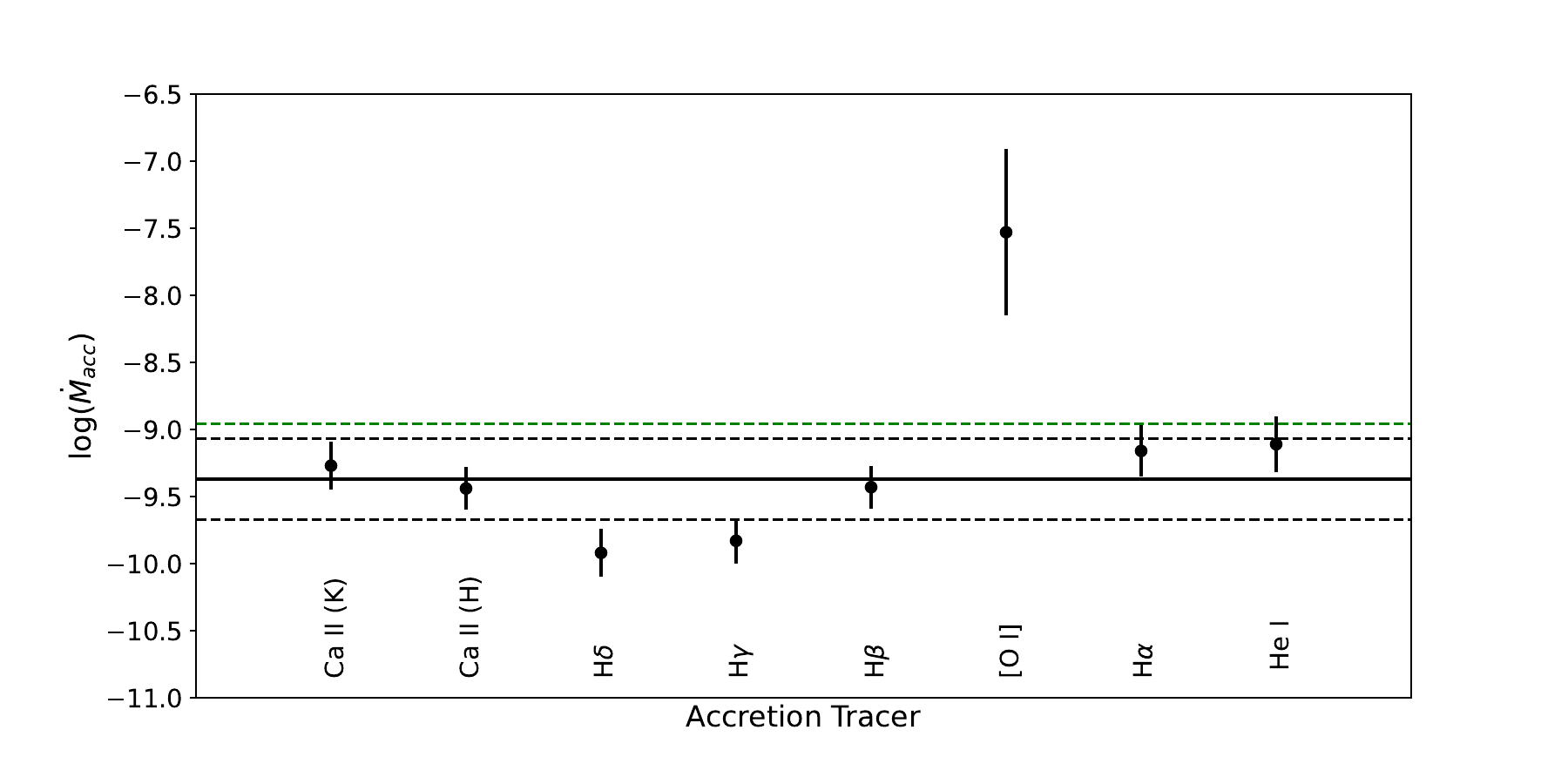}
   \caption{log($\dot M_{acc}$) measured using seven accretion-tracing lines and the [O~I]$\lambda$6300 jet emission.  The black solid and dashed lines are the measurements' mean and standard deviation.  The horizontal green dashed line marks the value estimated by \cite{Fang2023} also from a sample of accretion lines.}
  \label{ACC}     
\end{figure*}

The mass outflow rate as a fraction of the stellar mass accretion rate ($\dot M_{\rm out}$/$\dot M_{\rm acc}$) is called the outflow efficiency.  The mass accretion rate is estimated from the luminosity of lines known to be well correlated with accretion \citep{Alcala2017}.  Seven such lines are identified in the spectrum of 2M16075796: the Ca~II H and K lines, four Balmer lines (H$\delta$, H$\gamma$, H$\beta$ and H$\alpha$), and the He~I 5875~\AA\ line.  The accretion luminosity $L_{\rm acc}$ is related to $\dot M_{\rm acc}$ by
\begin{equation}
    \dot M_{\rm acc} = \frac{1.25 L_{\rm acc} R_{*}}{G M_{*}},
\end{equation}
while $L_{\rm acc}$ for each line is calculated from 
\begin{equation}
    \log L_{\rm acc} = A \log L_{\rm line} - B.
\end{equation}
The coefficients $A$ and $B$ come from \cite{Alcala2017}.  Before converting the line fluxes in Table 1 to $L_{\rm line}$ we correct for extinction using $A_V = 1.4$~mag from \cite{Fang2023}. The extinction correction was done using the using the extinction curve at Rv = 3.1 from \cite{Savage1979}. The accretion rate depends on the stellar mass and radius, $M_*=0.71$~\Msun\ and $R_*=0.59$~\Rsun\ \citep{Fang2023}.  The results shown in Figure~\ref{ACC} yield a mean $\log(\dot M_{\rm acc}/$\Msun${\rm yr}^{-1}) = -9.33$, lower by 0.37~dex than the value of $\log(\dot M_{acc}/$\Msun${\rm yr}^{-1}) = -8.96$ reported by \cite{Fang2023}.  Variability on this scale is typical of CTTSs \citep{Fischer2023, Costigan2012} and likely connected to the photometric variability of the source seen in Figure~1.  The error bars in Figure \ref{ACC} come from the uncertainties of the flux measurements and of the $A$ and $B$ coefficients. 

The mass loss rate in each outflow component is estimated from
\begin{equation}
  \dot M_{\rm out} = \frac{M_{\rm gas} V_{\rm jet}}{l_{\rm jet}},
\end{equation}
where $V_{\rm jet}$ is the jet velocity and $l_{\rm jet}$ the length of the emitting region.  Since all three components identified in the outflow have properties consistent with HVCs, we use the above equation to derive $\dot M_{\rm out}$ in the red and blue jets and the LVE.  The gas mass is derived from the \SIIa\ line luminosity using the equation in \cite{HartiganApJ1995} with the densities estimated above from the [S~II]$\lambda$6716/[S~II]$\lambda$6731 ratio.  The values of $\dot M_{\rm out}$ for the three outflow components in 2M160757 are reported in Table~\ref{Mout}. 

The values of $\dot M_{\rm out}$ listed in Table~\ref{Mout} are all in the ranges typical for CTT jets.  They are however larger than the mean value of $\dot M_{\rm acc}$ derived from the accretion tracers.  Given the system's inclination and the evidence for scattering in Figure~\ref{waterfall}, the accretion luminosity is likely underestimated with the accretion zone's emission scattered and obscured.  To test this, we use the \OIa\ jet emission to estimate $\dot M_{\rm acc}$ as described by \cite{Whelan2014}.  Due to the known correlation between mass outflow and accretion, \OIa\ is considered an indirect probe of accretion.  As it originates above the accretion zone it is less impacted by disk obscuration and has been used in the past to derive $\dot M_{\rm acc}$ for sources where the emission associated with the accretion is obscured \citep{Whelan2014AA}.  Taking the $A$ and $B$ coefficients for the \OIa\ line adopted in \cite{nisini2018} yields $\log(\dot M_{\rm out}/$\Msun yr$^{-1}) = -7.53$.  The outflow efficiencies of each of the components using $\dot M_{\rm acc}$ from the \OIa\ line are given in Table~\ref{Mout}.  All are $<10$~\%, which is typical for CTT jets \citep{HartiganApJ1995}.

\section{Discussion and Conclusions}
This spectroscopic study of 2M16075796 is an opportunity to investigate outflow activity in an older ($>5$~Myr) star-forming region.  The jet driven by 2M16075796 is typical of CTT jets in several respects.  The mass loss rate is within the range derived for CTTSs \citep{Nisini2023, nisini2018}.  If we surmise that $\dot M_{\rm acc}$ is underestimated from the accretion tracers and instead use the accretion rate derived from the \OIa\ line, the jet's two-sided efficiency is 8.6\%, also typical of CTT jets \cite{HartiganApJ1995, Ray2007}.  A further feature of the 2M16075796 jet commonly seen in CTTSs is the asymmetry between the blue and red lobes.  The red-shifted lobe is faster and denser while the blue-shifted lobe is brighter and has a higher mass flow rate.  This asymmetry does not extend to the positions of the furthest knots, however.  Likely the lack of symmetry between the inner red- and blue-shifted knots comes from the disk obscuring the inner part of the red-shifted flow.  A further important feature of this jet is the misalignment between the perpendicular to the disk and the jet PA estimated from the spectro-astrometry.  Two studies of the same dataset have put forward estimates of the disk major axis PA.  \cite{Barenfeld2017} put it at 0$^{\circ}$ and \cite{Hendler2020} 29$^{\circ}$, a significant difference.  The spectro-astrometric analysis puts the jet $<10^{\circ}$ off an east-west line.  This is more in agreement with the disk PA put forward by \cite{Barenfeld2017} but still suggests that the jet is not perpendicular to the disk.  Similar misalignments observed in other jets may come from bending or wiggling induced as the jet-launching star and its companion orbit one another \citep{Cox2013, Murphy2021, Whelan2023b}.  

While the jet of 2M16075796 is quite similar to those from other T~Tauri stars, the lower-velocity emission identified by \cite{Fang2023} is more of a puzzle.  One feature of the low-velocity emission that stands out is its smaller peak-to-continuum ratio when compared to the HVC \cite{Banzatti2019}.  More typical stars' have peak-to-continuum greater for the LVC than the HVC.  The large kinematic studies of the \OIa\ line in TTSs by \cite{Nisini2023, Banzatti2019, Fang2018, Simon2016} have some overlap in their source list and together represent 90 unique sources.  Of those, 50\% have a HVC and among this half of the combined sample we identify 8~sources (17\% of those with HVCs) where the line peak-to-continuum ratio of the HVC is equal to or greater than the LVC.  These, ordered by disk inclination angle increasing from $10^{\circ}$ to $62^{\circ}$ are SCrA~N, AS~353A, DO~Tau, IP~Tau, V2508 Oph, Sz~73, V853 Oph, and IQ~Tau.  This list omits sources identified as having edge-on disks (e.g.\ Sz~102) as the HVC and LVC would be indistinguishable in these cases.

The spread in inclination angles among the above sources suggests their lower apparent intensity of the LVC with respect to the HVC is not due to inclination.  The two sources most resembling 2M16075796 are V853 Oph and IQ~Tau.  These have the most-inclined disks, highlighting how the components become blended at the higher disk inclinations.  If the LVC is much less intense than the HVC, it will appear as a secondary peak or wing in these cases.  Further inspection of these sources yields no shared special properties apart from the shape of their \OIa\ line profiles.  They represent for example a range of stellar spectral types, masses, and $\dot M_{\rm acc}$.  Some of the sources may be binaries, including DO~Tau \citep{Erkal2021AAa} so possibly the peculiar line profiles come from combining the forbidden line emission from more than one outflow-driving source.  High-resolution imaging may help further understand the outflows in these systems. 

While the peak velocity of the LVE in 2M16075796 straddles the boundary used to separate a HVC from a LVC, all other properties of the LVE show it to be compatible with a HVC.  The \NII\ line is not detected in LVCs but here the LVE is detected in \NII.  The spectro-astrometric analysis reveals a gradient typical of a jet and opposite to that which has been associated with a wind \citep{Takami2001, Whelan2021}.  The diagnostic analysis indicates a density and ionisation fraction for the LVE typical of a HVC with $x_e$ in particular far higher than measured to date in LVCs.  The only property of the LVE which has been seen before in LVCs is the increase in peak velocity with decreasing critical density.  This has been interpreted as tracing an accelerating flow with a density dependence, so that material higher up in the flow moves faster but is lower in density and so brighter in the FELs with lower critical densities.  As is clear from Figure~\ref{Lines_comp}, the HVC shows no such density dependence.  This is compatible with emission dominated by shocks.  Figure~\ref{PVs_B} shows forbidden Ca emission in both jets but an absence of emission in the velocity range of the LVE.  This is compatible with shocks strong enough and temperatures high enough to destroy the dust grains and release calcium into the gas. 

\begin{figure*}
\centering
\includegraphics[width=15cm, trim= 0cm 5cm 0cm 0cm, clip=true]{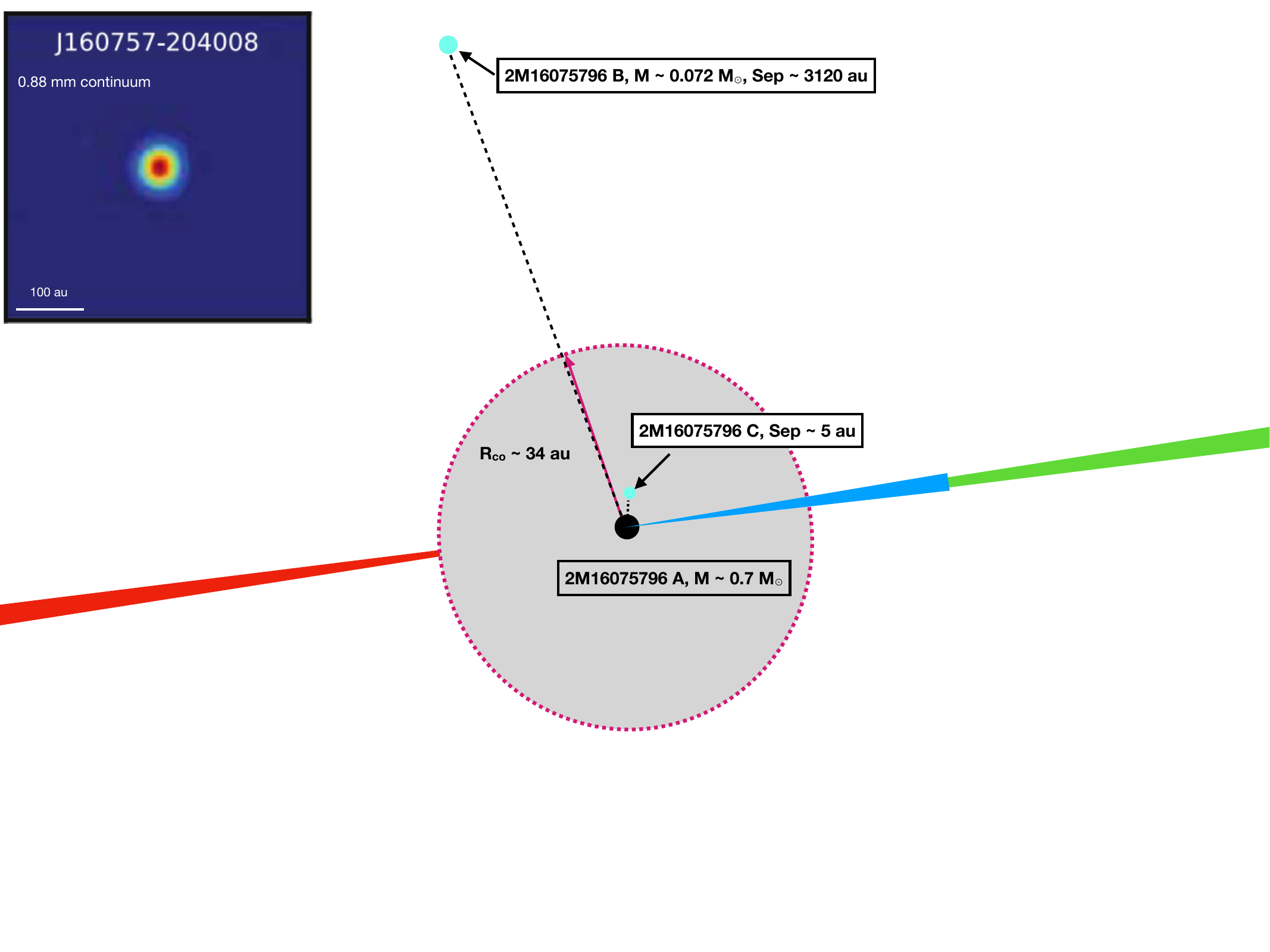}
   \caption{{\bf Sketch of the 2M16075796 system with the ALMA 0.88~mm continuum image inset to provide context for the depicted morphology of the 2M16075796 accretion disk.} The separation of the outer companion is not to scale. The kinematics, spectro-astrometry, and line ratios support the presence of two blue-shifted jets with a radial velocity ratio 0.36 in \SIIa.  These are shown overlapping in blue and green.  The two jets' PAs differ by less than the 5$^{\circ}$ uncertainty of the spectro-astrometric measurements.  }
  \label{cartoon}     
\end{figure*}

\cite{Fang2023} found that only five sources in their Upper Sco sample of 115 CTTSs had jets.  This makes 2M16075796 unique in this sample. 2M16075796 also stands out from the four other CTTSs with jets as the only one classified by \cite{Cody2017} as a burster.  Two of the other sources have also been associated with companions \citep{Barenfeld2019}.  Figure~\ref{cartoon} schematically illustrates the structure we propose for the system.  One possibility is that the close companion induces accretion bursts which are sustaining the jet long after the age which one would not expect to find jets \citep{Lodato2004, Whelan2010, Cody2017, Hales2018}.  Looking at the blue-shifted flow we see knots with dynamical ages of 5.2 and 0.87~years, suggesting a time interval of $\sim$ 4~years between ejections. It is not fully clear how the time between knots might relate to the orbital period of the companion but a period of 4~years would suggest a companion separation closer to 2~au than the 4.6~au reported by \cite{Barenfeld2019}. More study is needed in this area. The LVE emission could be a low-velocity jet driven by the companion, with both outflows being launched from the circumbinary disk \citep{Whelan2010, Lynch2020a, Lynch2020b}. The peak velocities measured for the LVE are within the range recorded for the FELs tracing BD jets \citep{Whelan2014b}. In this scenario any low velocity emission from a MHD wind would be blended with the low velocity jet making it difficult to isolate the MHD wind component to the FEL region.

The study of 2M16075796 presented here highlights the complexity which can be present in the individual objects making up the population-level kinematic studies of the origin of the FEL LVCs.  Such studies reveal many objects worthy of detailed follow-up.  Particularly valuable would be to determine whether the other sources where the LVC occurs as a secondary peak resemble 2M16075796 in showing evidence of binarity or a second, lower-velocity jet.  Suitable follow-up approaches include spectro-astrometry and high-angular-resolution integral-field spectroscopy. Future spectro-astrometric studies would benefit from higher spectral resolution to better separate high and low velocity emission. Adaptive optics assisted integral field spectroscopy typically achieves a spatial resolution of 5 to 10~au for the nearest star forming regions \citep{Whelan2010, Fang2023N, Flores2023}. The European Southern Observatory's Multi Unit Spectroscopic Explorer (MUSE) in narrow field mode (NFM) achieves this for the FELs investigated in this work. In the case of 2M16075796 and remembering that spectro-astrometry records the minimum spatial extent of the outflow components, MUSE NFM observations with a spatial resolution $<$ 10~au could be used to produce spectro-images of the jet and LVE which could then be used to better understand the morphology, PA and collimation of the two outflow components \citep{Birney2024}. While these observations are unlikely to directly resolve the close companion, the jet axis position could be mapped to look for signatures of the companion  \citep{Murphy2021}. The Keck OSIRIS or ESO ERIS instruments would offer similar observations in the near-infrared where lines such as the [Fe II]1.644 micron lines are strong jet tracers. ERIS has an advantage over other integral field spectrographs as it has a spectral resolution a factor of 3 better, meaning that it would more effectively separate the two outflow components

Another approach to understanding such outflows is that of the proposed Hyperion space telescope, which would use spectroscopy of molecular hydrogen's many far-UV fluorescent emission lines to directly measure the total molecular masses and speeds of disk winds \citep{2022Hamden}.  The flows from the inner and outer disks would be distinguished by their different fluorescent spectra resulting from different illuminating radiation fields.  Combining the far-UV measurements with spectro-astrometric flow heights from ground-based telescopes like those we report here would make it possible to determine the mass outflow rates via eq.~(4) without the uncertainties inherent in extrapolating the wind mass from trace species such as oxygen and sulfur.  Comparison with the stellar accretion rates would then enable gauging the extent to which the winds drive the disks' evolution \citep{2022Hasegawa}.  Separating the molecular hydrogen fluorescent lines requires high spectral resolution and high signal-to-noise ratios, factors that motivate Hyperion's design.  The mission was proposed to NASA's Astrophysics Medium Explorer program in 2021 and rated in the highest category but not selected.  A similar concept will be re-proposed in a future call.

\acknowledgments
We are grateful to John O'Meara and the Keck Telescopes team for carrying out the observations and to Ann Marie Cody for discussion of burster stars.  NJT's efforts were funded by NASA Exoplanets Research Program award 17-XRP17\_2-0081.  The work was carried out in part at the Jet Propulsion Laboratory, California Institute of Technology, under contract 80NM00018D0004 with NASA.

\begin{appendix}
    
\begin{figure*}
\centering
\includegraphics[width=11.8cm, trim= 0cm 0cm 0cm 0cm, clip=true]{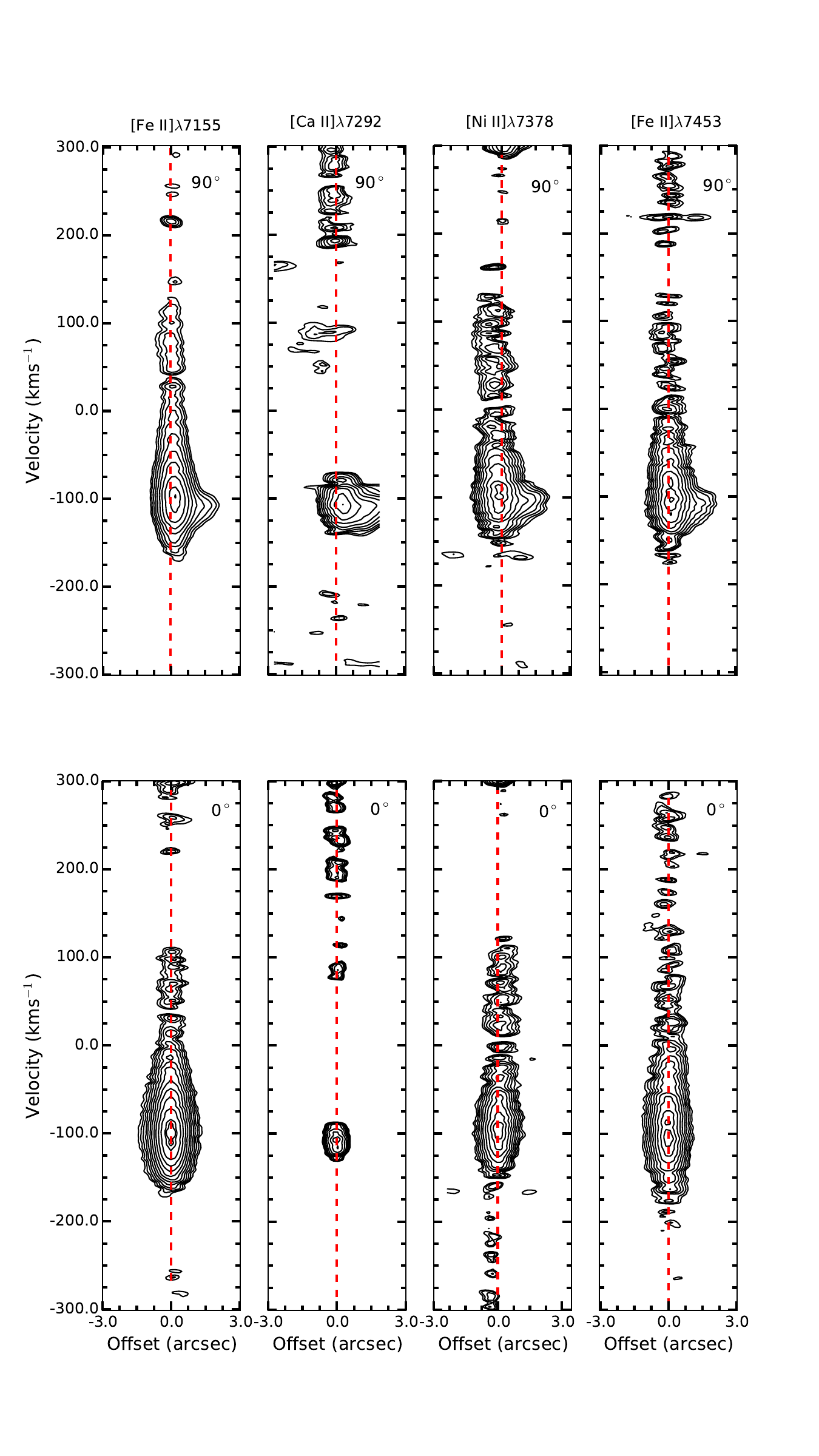}
   \caption{Continuum subtracted position velocity diagrams of key FELs. Contours begin at the 3 times the rms noise and increase with an interval of $\sqrt{2}$. The top row is from the spectrum at the slit PA of 90$^{\circ}$ and the bottom row from the spectrum at the slit PA of 0$^{\circ}$. Both the blue and red-shifted high velocity emission at 90$^{\circ}$ is extended while at 0$^{\circ}$ it is compact by eye. This reveals that the jet traced by the HVC has a PA close to 90$^{\circ}$.}
  \label{PVs_B}     
\end{figure*}

\begin{figure*}
\centering
\includegraphics[width=20cm, trim= 2cm 1cm 2cm 2cm, clip=true]{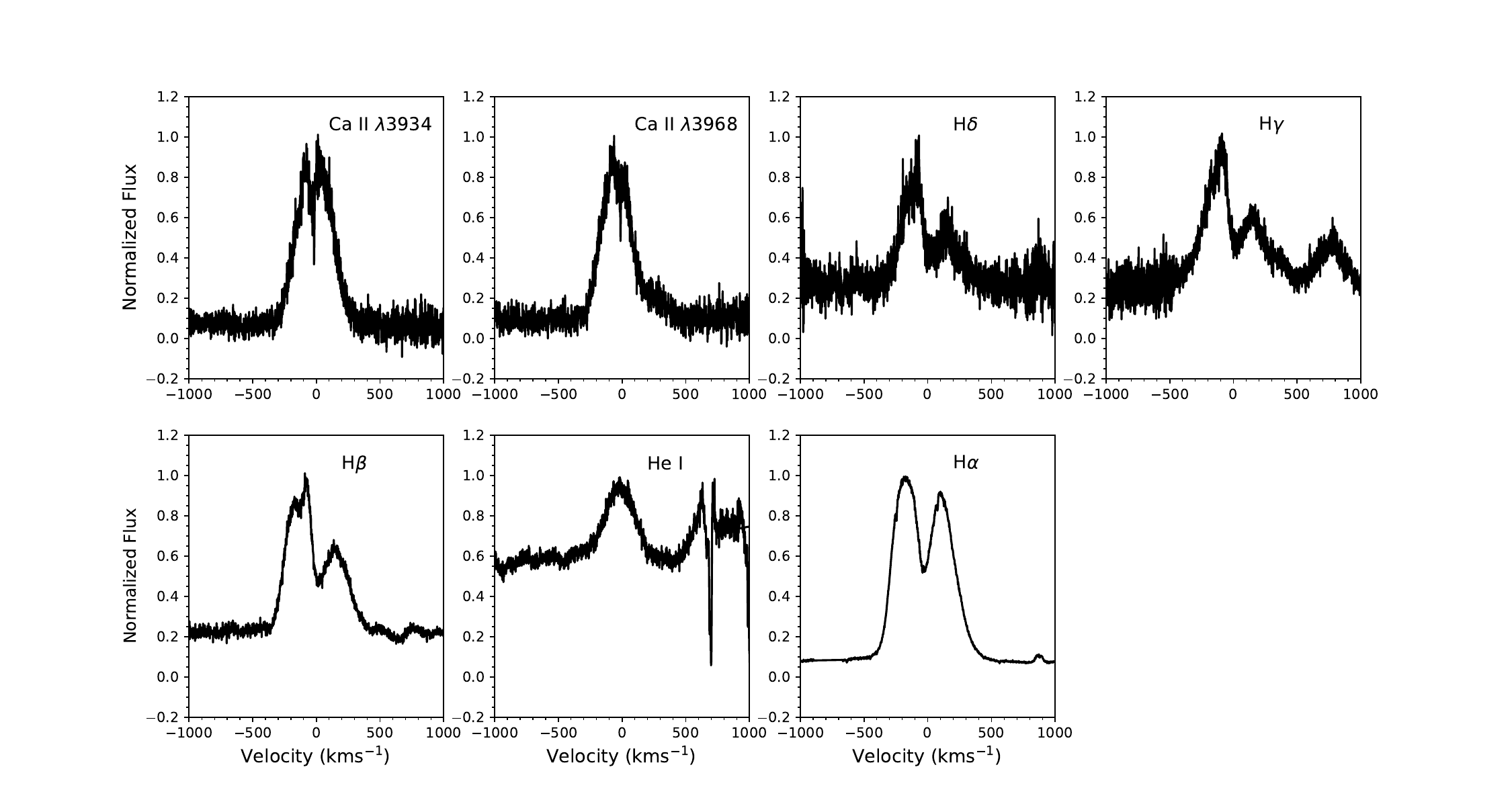}
   \caption{Line profiles of the accretion tracers listed in Table \ref{fluxes} and used in the calculation of the mass accretion rate.}
  \label{Acc_lines}     
\end{figure*}

\begin{figure*}
\centering
\includegraphics[width=22cm, trim= 5cm 0cm 0cm 0cm, clip=true]{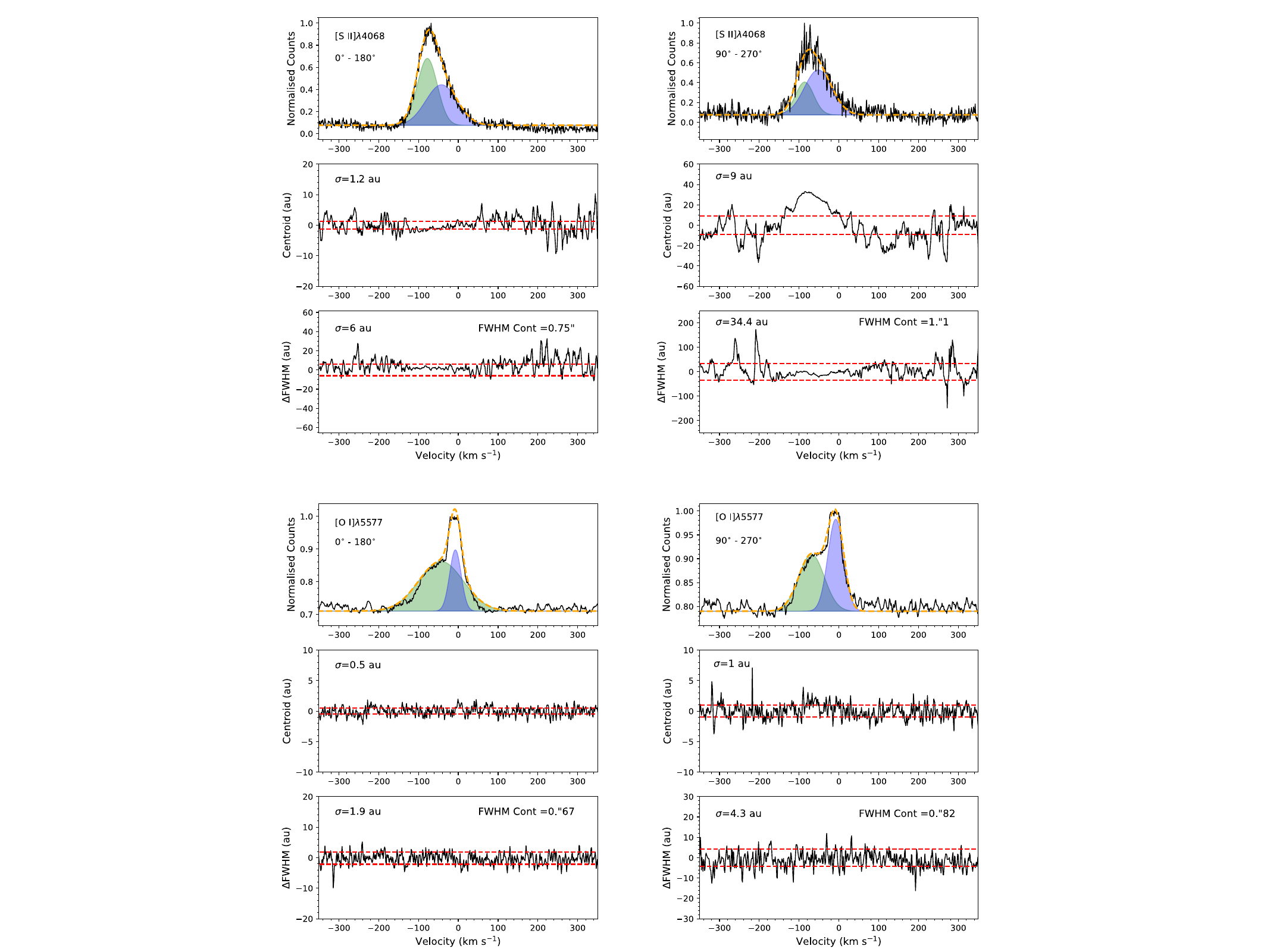}
   \caption{Spectro-astrometric analysis of the \SIIb\ and \OIb\ lines.  Results for the N-S and E-W slit PAs are presented in the left and right columns respectively.  The \SIIb\ line is in the upper two quadrants and the \OIb\ line in the lower two.  The three panels within each quadrant are from top to bottom the line emission normalized to the line peak, the centroid, and the FWHM.  The fitted Gaussian kinematic components are overplotted.  The centroids have not been corrected for continuum contamination.  Low signal-to-noise ratios in the continuum at the wavelength of the \SIIb\ line and in the line emission make it difficult to extract a clear spectro-astrometric signal for the \SIIb\ line.  However, there is a signal consistent with the jet.  The \OIb\ emission is compact.}
  \label{SA2}     
\end{figure*}

\begin{figure*}
\centering
\includegraphics[width=22cm, trim= 5cm 0cm 0cm 0cm, clip=true]{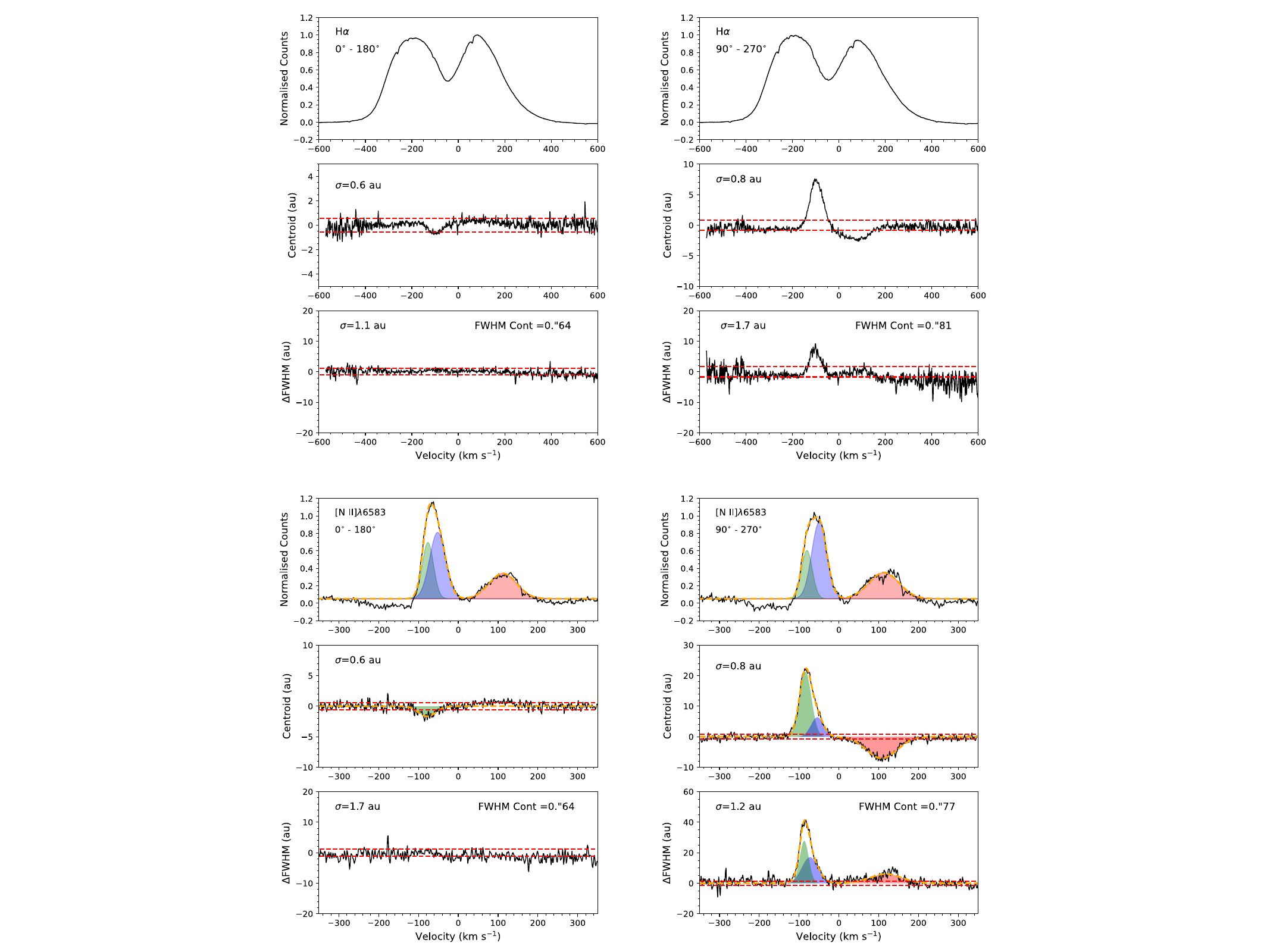}
   \caption{Spectro-astrometric analysis of the H$\alpha$ and [N~II]$\lambda$6583 lines.  Results for the N-S and E-W slit PAs are presented in the left and right columns, respectively.  The H$\alpha$ line is in the upper two quadrants and the [N~II]$\lambda$6583 line in the lower two.  The three panels within each quadrant are from top to bottom the line emission normalized to the line peak, the centroid, and the FWHM.  The centroids have not been corrected for continuum contamination.  For the [N~II]$\lambda$6583 line, the Gaussian components fitted to the line profiles, centroids, and $\Delta$FWHM are overplotted.  In both lines, a shift in the centroid at the velocity of the jet to $\sim$2~au is measured in the N-S spectrum, indicating the jet's alignment is not perfectly E-W.}
  \label{SA3}     
\end{figure*}

\end{appendix}


\begin{thebibliography}{}
\expandafter\ifx\csname natexlab\endcsname\relax\def\natexlab#1{#1}\fi
\providecommand{\url}[1]{\href{#1}{#1}}
\providecommand{\dodoi}[1]{doi:~\href{http://doi.org/#1}{\nolinkurl{#1}}}
\providecommand{\doeprint}[1]{\href{http://ascl.net/#1}{\nolinkurl{http://ascl.net/#1}}}
\providecommand{\doarXiv}[1]{\href{https://arxiv.org/abs/#1}{\nolinkurl{https://arxiv.org/abs/#1}}}

\bibitem[{{Alcal{\'a}} {et~al.}(2017){Alcal{\'a}}, {Manara}, {Natta}, {Frasca},
  {Testi}, {Nisini}, {Stelzer}, {Williams}, {Antoniucci}, {Biazzo}, {Covino},
  {Esposito}, {Getman}, \& {Rigliaco}}]{Alcala2017}
{Alcal{\'a}}, J.~M., {Manara}, C.~F., {Natta}, A., {et~al.} 2017, \aap, 600,
  A20, \dodoi{10.1051/0004-6361/201629929}

\bibitem[{{Bailey}(1998)}]{Bailey1998}
{Bailey}, J. 1998, \mnras, 301, 161, \dodoi{10.1046/j.1365-8711.1998.02010.x}

\bibitem[{{Banzatti} {et~al.}(2019){Banzatti}, {Pascucci}, {Edwards}, {Fang},
  {Gorti}, \& {Flock}}]{Banzatti2019}
{Banzatti}, A., {Pascucci}, I., {Edwards}, S., {et~al.} 2019, \apj, 870, 76,
  \dodoi{10.3847/1538-4357/aaf1aa}

\bibitem[{{Barenfeld} {et~al.}(2016){Barenfeld}, {Carpenter}, {Ricci}, \&
  {Isella}}]{Barenfield2016}
{Barenfeld}, S.~A., {Carpenter}, J.~M., {Ricci}, L., \& {Isella}, A. 2016,
  \apj, 827, 142, \dodoi{10.3847/0004-637X/827/2/142}

\bibitem[{{Barenfeld} {et~al.}(2017){Barenfeld}, {Carpenter}, {Sargent},
  {Isella}, \& {Ricci}}]{Barenfeld2017}
{Barenfeld}, S.~A., {Carpenter}, J.~M., {Sargent}, A.~I., {Isella}, A., \&
  {Ricci}, L. 2017, \apj, 851, 85, \dodoi{10.3847/1538-4357/aa989d}

\bibitem[{{Barenfeld} {et~al.}(2019){Barenfeld}, {Carpenter}, {Sargent},
  {Rizzuto}, {Kraus}, {Meshkat}, {Akeson}, {Jensen}, \&
  {Hinkley}}]{Barenfeld2019}
{Barenfeld}, S.~A., {Carpenter}, J.~M., {Sargent}, A.~I., {et~al.} 2019, \apj,
  878, 45, \dodoi{10.3847/1538-4357/ab1e50}

\bibitem[{{Birney} \& {Whelan}(2024)}]{Birney2024}
{Birney}, M., \& {Whelan}, E. 2024, \aap

\bibitem[{{Brannigan} {et~al.}(2006){Brannigan}, {Takami}, {Chrysostomou}, \&
  {Bailey}}]{Brannigan2006}
{Brannigan}, E., {Takami}, M., {Chrysostomou}, A., \& {Bailey}, J. 2006,
  \mnras, 367, 315, \dodoi{10.1111/j.1365-2966.2005.09942.x}

\bibitem[{{Cazzoletti} {et~al.}(2019){Cazzoletti}, {Manara}, {Liu}, {van
  Dishoeck}, {Facchini}, {Alcal{\`a}}, {Ansdell}, {Testi}, {Williams},
  {Carrasco-Gonz{\'a}lez}, {Dong}, {Forbrich}, {Fukagawa}, {Galv{\'a}n-Madrid},
  {Hirano}, {Hogerheijde}, {Hasegawa}, {Muto}, {Pinilla}, {Takami}, {Tamura},
  {Tazzari}, \& {Wisniewski}}]{Cazzoletti2019}
{Cazzoletti}, P., {Manara}, C.~F., {Liu}, H.~B., {et~al.} 2019, \aap, 626, A11,
  \dodoi{10.1051/0004-6361/201935273}

\bibitem[{{Cody} {et~al.}(2017){Cody}, {Hillenbrand}, {David}, {Carpenter},
  {Everett}, \& {Howell}}]{Cody2017}
{Cody}, A.~M., {Hillenbrand}, L.~A., {David}, T.~J., {et~al.} 2017, \apj, 836,
  41, \dodoi{10.3847/1538-4357/836/1/41}

\bibitem[{{Comer{\'o}n} \& {Fern{\'a}ndez}(2011)}]{Comeron2011}
{Comer{\'o}n}, F., \& {Fern{\'a}ndez}, M. 2011, \aap, 528, A99,
  \dodoi{10.1051/0004-6361/201015983}

\bibitem[{{Costigan} {et~al.}(2012){Costigan}, {Scholz}, {Stelzer}, {Ray},
  {Vink}, \& {Mohanty}}]{Costigan2012}
{Costigan}, G., {Scholz}, A., {Stelzer}, B., {et~al.} 2012, \mnras, 427, 1344,
  \dodoi{10.1111/j.1365-2966.2012.22008.x}

\bibitem[{{Cox} {et~al.}(2013){Cox}, {Grady}, {Hammel}, {Hornbeck}, {Russell},
  {Sitko}, \& {Woodgate}}]{Cox2013}
{Cox}, A.~W., {Grady}, C.~A., {Hammel}, H.~B., {et~al.} 2013, \apj, 762, 40,
  \dodoi{10.1088/0004-637X/762/1/40}

\bibitem[{{Dougados} {et~al.}(2000){Dougados}, {Cabrit}, {Lavalley}, \&
  {M{\'e}nard}}]{Dougados2000}
{Dougados}, C., {Cabrit}, S., {Lavalley}, C., \& {M{\'e}nard}, F. 2000, \aap,
  357, L61

\bibitem[{{Erkal} {et~al.}(2021){Erkal}, {Dougados}, {Coffey}, {Cabrit},
  {Bacciotti}, {Garcia-Lopez}, {Fedele}, \& {Chrysostomou}}]{Erkal2021AAa}
{Erkal}, J., {Dougados}, C., {Coffey}, D., {et~al.} 2021, \aap, 650, A46,
  \dodoi{10.1051/0004-6361/202038977}

\bibitem[{{Espaillat} {et~al.}(2021){Espaillat}, {Robinson}, {Romanova},
  {Thanathibodee}, {Wendeborn}, {Calvet}, {Reynolds}, \&
  {Muzerolle}}]{Espaillat2021}
{Espaillat}, C.~C., {Robinson}, C.~E., {Romanova}, M.~M., {et~al.} 2021, \nat,
  597, 41, \dodoi{10.1038/s41586-021-03751-5}

\bibitem[{{Fang} {et~al.}(2018){Fang}, {Pascucci}, {Edwards}, {Gorti},
  {Banzatti}, {Flock}, {Hartigan}, {Herczeg}, \& {Dupree}}]{Fang2018}
{Fang}, M., {Pascucci}, I., {Edwards}, S., {et al.} 2018{\natexlab{a}}, \apj, 868, 28, \dodoi{10.3847/1538-4357/aae780}

\bibitem[{{Fang} {et~al.}(2023{\natexlab{a}}){Fang}, {Pascucci}, {Edwards},
  {Gorti}, {Hillenbrand}, \& {Carpenter}}]{Fang2023}
{Fang}, M., {Pascucci}, I., {Edwards}, S., {et~al.} 2023{\natexlab{a}}, \apj,
  945, 112, \dodoi{10.3847/1538-4357/acb2c9}

  \bibitem[{{Fang} {et~al.}(2023{\natexlab{b}}){Fang}, {Wang}, {Herczeg},
  {Hashimoto}, {Xu}, {Nemer}, {Pascucci}, {Haffert}, \& {Aoyama}}]{Fang2023N}
{Fang}, M., {Wang}, L., {Herczeg}, G.~J., {et~al.} 2023{\natexlab{b}}, Nature
  Astronomy, 7, 905, \dodoi{10.1038/s41550-023-02004-x}






\bibitem[{{Fedriani} {et~al.}(2018){Fedriani}, {Caratti o Garatti}, {Coffey},
  {Garcia Lopez}, {Kraus}, {Weigelt}, {Stecklum}, {Ray}, \&
  {Walmsley}}]{Fedriani2018}
{Fedriani}, R., {Caratti o Garatti}, A., {Coffey}, D., {et~al.} 2018, \aap,
  616, A126, \dodoi{10.1051/0004-6361/201732180}

\bibitem[{{Fischer} {et~al.}(2023){Fischer}, {Hillenbrand}, {Herczeg},
  {Johnstone}, {Kospal}, \& {Dunham}}]{Fischer2023}
{Fischer}, W.~J., {Hillenbrand}, L.~A., {Herczeg}, G.~J., {et~al.} 2023, in
  Astronomical Society of the Pacific Conference Series, Vol. 534, Protostars
  and Planets VII, ed. S.~{Inutsuka}, Y.~{Aikawa}, T.~{Muto}, K.~{Tomida}, \&
  M.~{Tamura}, 355, \dodoi{10.48550/arXiv.2203.11257}

\bibitem[{{Flores-Rivera} {et~al.}(2023){Flores-Rivera}, {Flock}, {Kurtovic},
  {Husemann}, {Banzatti}, {Ringqvist}, {Kamann}, {M{\"u}ller}, {Fendt},
  {Garc{\'\i}a Lopez}, {Marleau}, {Henning}, {Carrasco-Gonz{\'a}lez}, {van
  Boekel}, {Keppler}, {Launhardt}, \& {Aoyama}}]{Flores2023}
{Flores-Rivera}, L., {Flock}, M., {Kurtovic}, N.~T., {et~al.} 2023, \aap, 670,
  A126, \dodoi{10.1051/0004-6361/202141664}

\bibitem[{{Frank} {et~al.}(2014){Frank}, {Ray}, {Cabrit}, {Hartigan}, {Arce},
  {Bacciotti}, {Bally}, {Benisty}, {Eisl{\"o}ffel}, {G{\"u}del}, {Lebedev},
  {Nisini}, \& {Raga}}]{Frank2014}
{Frank}, A., {Ray}, T.~P., {Cabrit}, S., {et~al.} 2014, in Protostars and
  Planets VI, ed. H.~{Beuther}, R.~S. {Klessen}, C.~P. {Dullemond}, \&
  T.~{Henning}, 451--474, \dodoi{10.2458/azu\_uapress\_9780816531240-ch020}

\bibitem[{{Gangi} {et~al.}(2023){Gangi}, {Nisini}, {Manara}, {France},
  {Antoniucci}, {Biazzo}, {Giannini}, {Herczeg}, {Alcal{\'a}}, {Frasca},
  {Mauc{\'o}}, {Campbell-White}, {Siwak}, {Venuti}, {Schneider},
  {K{\'o}sp{\'a}l}, {Caratti o Garatti}, {Fiorellino}, {Rigliaco}, \&
  {Yadav}}]{Gangi2023}
{Gangi}, M., {Nisini}, B., {Manara}, C.~F., {et~al.} 2023, \aap, 675, A153,
  \dodoi{10.1051/0004-6361/202346543}

\bibitem[{{Giannini} {et~al.}(2019){Giannini}, {Nisini}, {Antoniucci},
  {Biazzo}, {Alcal{\'a}}, {Bacciotti}, {Fedele}, {Frasca}, {Harutyunyan},
  {Munari}, {Rigliaco}, \& {Vitali}}]{Giannini2019}
{Giannini}, T., {Nisini}, B., {Antoniucci}, S., {et~al.} 2019, \aap, 631, A44,
  \dodoi{10.1051/0004-6361/201936085}

\bibitem[{{Gnerucci} {et~al.}(2010){Gnerucci}, {Marconi}, {Capetti}, {Axon}, \&
  {Robinson}}]{Gnerucci2010}
{Gnerucci}, A., {Marconi}, A., {Capetti}, A., {Axon}, D.~J., \& {Robinson}, A.
  2010, \aap, 511, A19, \dodoi{10.1051/0004-6361/200912530}

\bibitem[{{Hales} {et~al.}(2018){Hales}, {P{\'e}rez}, {Saito}, {Pinte}, {Knee},
  {de Gregorio-Monsalvo}, {Dent}, {L{\'o}pez}, {Plunkett}, {Cort{\'e}s},
  {Corder}, \& {Cieza}}]{Hales2018}
{Hales}, A.~S., {P{\'e}rez}, S., {Saito}, M., {et~al.} 2018, \apj, 859, 111,
  \dodoi{10.3847/1538-4357/aac018}

\bibitem[{{Hamden} {et~al.}(2022){Hamden}, {Schiminovich}, {Nikzad}, {Turner},
  {Burkhart}, {Haworth}, {Hoadley}, {Serena Kim}, {Bialy}, {Bryden}, {Chung},
  {Imara}, {Kennicutt}, {Pineda}, {Kong}, {Hasegawa}, {Pascucci}, {Godard},
  {Krumholz}, {Lee}, {Seifried}, {Sternberg}, {Walch}, {Smith}, {Unwin},
  {Luthman}, {Kiessling}, {McGuire}, {Rais-Zadeh}, {Hoenk}, {Pavlak}, {Vargas},
  \& {Kim}}]{2022Hamden}
{Hamden}, E.~T., {Schiminovich}, D., {Nikzad}, S., {et~al.} 2022, Journal of
  Astronomical Telescopes, Instruments, and Systems, 8, 044008,
  \dodoi{10.1117/1.JATIS.8.4.044008}

\bibitem[{{Hartigan} {et~al.}(1995){Hartigan}, {Edwards}, \&
  {Ghandour}}]{HartiganApJ1995}
{Hartigan}, P., {Edwards}, S., \& {Ghandour}, L. 1995, \apj, 452, 736,
  \dodoi{10.1086/176344}

\bibitem[{{Hartigan} {et~al.}(1989){Hartigan}, {Hartmann}, {Kenyon}, {Hewett},
  \& {Stauffer}}]{Hartigan1989}
{Hartigan}, P., {Hartmann}, L., {Kenyon}, S., {Hewett}, R., \& {Stauffer}, J.
  1989, \apjs, 70, 899, \dodoi{10.1086/191361}

\bibitem[{{Hasegawa} {et~al.}(2022){Hasegawa}, {Haworth}, {Hoadley}, {Kim},
  {Goto}, {Juzikenaite}, {Turner}, {Pascucci}, \& {Hamden}}]{2022Hasegawa}
{Hasegawa}, Y., {Haworth}, T.~J., {Hoadley}, K., {et~al.} 2022, \apjl, 926,
  L23, \dodoi{10.3847/2041-8213/ac50aa}

\bibitem[{{Hendler} {et~al.}(2020){Hendler}, {Pascucci}, {Pinilla}, {Tazzari},
  {Carpenter}, {Malhotra}, \& {Testi}}]{Hendler2020}
{Hendler}, N., {Pascucci}, I., {Pinilla}, P., {et~al.} 2020, \apj, 895, 126,
  \dodoi{10.3847/1538-4357/ab70ba}

\bibitem[{{Hirth} {et~al.}(1997){Hirth}, {Mundt}, \& {Solf}}]{Hirth1997}
{Hirth}, G.~A., {Mundt}, R., \& {Solf}, J. 1997, \aaps, 126, 437,
  \dodoi{10.1051/aas:1997275}

\bibitem[{{Kraus} \& {Hillenbrand}(2007)}]{Kraus2007a}
{Kraus}, A.~L., \& {Hillenbrand}, L.~A. 2007, \apj, 662, 413,
  \dodoi{10.1086/516835}

\bibitem[{{Kraus} \& {Hillenbrand}(2009)}]{Kraus2009}
{Kraus}, A.~L. \& {Hillenbrand}, L.~A.\ 2009, \apj, 704, 531, \dodoi{10.1088/0004-637X/703/2/1511}



\bibitem[{{Kwan} \& {Tademaru}(1988)}]{Kwan1988}
{Kwan}, J., \& {Tademaru}, E. 1988, \apjl, 332, L41, \dodoi{10.1086/185262}

\bibitem[{{Lodato} \& {Clarke}(2004)}]{Lodato2004}
{Lodato}, G., \& {Clarke}, C.~J. 2004, \mnras, 353, 841,
  \dodoi{10.1111/j.1365-2966.2004.08112.x}

\bibitem[{{Luhman} \& {Esplin}(2020)}]{Luhman2020}
{Luhman}, K.~L., \& {Esplin}, T.~L. 2020, \aj, 160, 44,
  \dodoi{10.3847/1538-3881/ab9599}

\bibitem[{{Lynch} \& {Smith}(2020)}]{Lynch2020b}
{Lynch}, C. J.~R., \& {Smith}, M.~D. 2020, \mnras, 494, 2299,
  \dodoi{10.1093/mnras/staa860}

\bibitem[{{Lynch} {et~al.}(2020){Lynch}, {Smith}, \& {Glover}}]{Lynch2020a}
{Lynch}, C. J.~R., {Smith}, M.~D., \& {Glover}, S. C.~O. 2020, \mnras, 491,
  3082, \dodoi{10.1093/mnras/stz2985}

\bibitem[{{McGroarty} \& {Ray}(2004)}]{McGroarty2004}
{McGroarty}, F., \& {Ray}, T.~P. 2004, \aap, 420, 975,
  \dodoi{10.1051/0004-6361:20041124}

\bibitem[{{Miret-Roig} {et~al.}(2022){Miret-Roig}, {Galli}, {Olivares}, {Bouy},
  {Alves}, \& {Barrado}}]{Miret-Roig2022}
{Miret-Roig}, N., {Galli}, P.~A.~B., {Olivares}, J., {et~al.} 2022, \aap, 667,
  A163, \dodoi{10.1051/0004-6361/202244709}

\bibitem[{{Murphy} {et~al.}(2021){Murphy}, {Dougados}, {Whelan}, {Bacciotti},
  {Coffey}, {Comer{\'o}n}, {Eisl{\"o}ffel}, \& {Ray}}]{Murphy2021}
{Murphy}, A., {Dougados}, C., {Whelan}, E.~T., {et~al.} 2021, \aap, 652, A119,
  \dodoi{10.1051/0004-6361/202141315}

\bibitem[{{Nisini} {et~al.}(2018){Nisini}, {Antoniucci}, {Alcal{\'a}},
  {Giannini}, {Manara}, {Natta}, {Fedele}, \& {Biazzo}}]{nisini2018}
{Nisini}, B., {Antoniucci}, S., {Alcal{\'a}}, J.~M., {et~al.} 2018, \aap, 609,
  A87, \dodoi{10.1051/0004-6361/201730834}

\bibitem[{{Nisini} {et~al.}(2024){Nisini}, {Gangi}, {Giannini}, {Antoniucci},
  {Biazzo}, {Frasca}, {Alcal{\'a}}, {Manara}, \& {Weber}}]{Nisini2023}
{Nisini}, B., {Gangi}, M., {Giannini}, T., {et~al.} 2024, \aap, 683, A116,
  \dodoi{10.1051/0004-6361/202346742}

\bibitem[{{Pascucci} {et~al.}(2022){Pascucci}, {Cabrit}, {Edwards}, {Gorti},
  {Gressel}, \& {Suzuki}}]{Pascucci2022}
{Pascucci}, I., {Cabrit}, S., {Edwards}, S., {et~al.} 2022, arXiv e-prints,
  arXiv:2203.10068, \dodoi{10.48550/arXiv.2203.10068}

\bibitem[{{Pascucci} {et~al.}(2020){Pascucci}, {Banzatti}, {Gorti}, {Fang},
  {Pontoppidan}, {Alexander}, {Ballabio}, {Edwards}, {Salyk}, {Sacco},
  {Flaccomio}, {Blake}, {Carmona}, {Hall}, {Kamp}, {K{\"a}ufl}, {Meeus},
  {Meyer}, {Pauly}, {Steendam}, \& {Sterzik}}]{Pascucci2020}
{Pascucci}, I., {Banzatti}, A., {Gorti}, U., {et~al.} 2020, \apj, 903, 78,
  \dodoi{10.3847/1538-4357/abba3c}

\bibitem[{{Pontoppidan} {et~al.}(2011){Pontoppidan}, {Blake}, \&
  {Smette}}]{Ponto2011}
{Pontoppidan}, K.~M., {Blake}, G.~A., \& {Smette}, A. 2011, \apj, 733, 84,
  \dodoi{10.1088/0004-637X/733/2/84}

\bibitem[{{Porter} {et~al.}(2004){Porter}, {Oudmaijer}, \&
  {Baines}}]{Porter2004}
{Porter}, J.~M., {Oudmaijer}, R.~D., \& {Baines}, D. 2004, \aap, 428, 327,
  \dodoi{10.1051/0004-6361:20035686}

\bibitem[{{Ratzenb{\"o}ck} {et~al.}(2023{\natexlab{a}}){Ratzenb{\"o}ck},
  {Gro{\ss}schedl}, {M{\"o}ller}, {Alves}, {Bomze}, \&
  {Meingast}}]{Ratzenbock2023a}
{Ratzenb{\"o}ck}, S., {Gro{\ss}schedl}, J.~E., {M{\"o}ller}, T., {et~al.}
  2023{\natexlab{a}}, \aap, 677, A59, \dodoi{10.1051/0004-6361/202243690}

\bibitem[{{Ratzenb{\"o}ck} {et~al.}(2023{\natexlab{b}}){Ratzenb{\"o}ck},
  {Gro{\ss}schedl}, {Alves}, {Miret-Roig}, {Bomze}, {Forbes}, {Goodman},
  {Hacar}, {Lin}, {Meingast}, {M{\"o}ller}, {Piecka}, {Posch}, {Rottensteiner},
  {Swiggum}, \& {Zucker}}]{Ratzenbock2023b}
{Ratzenb{\"o}ck}, S., {Gro{\ss}schedl}, J.~E., {Alves}, J., {et~al.}
  2023{\natexlab{b}}, \aap, 678, A71, \dodoi{10.1051/0004-6361/202346901}

\bibitem[{{Ray} {et~al.}(2007){Ray}, {Dougados}, {Bacciotti}, {Eisl{\"o}ffel},
  \& {Chrysostomou}}]{Ray2007}
{Ray}, T., {Dougados}, C., {Bacciotti}, F., {Eisl{\"o}ffel}, J., \&
  {Chrysostomou}, A. 2007, in Protostars and Planets V, ed. B.~{Reipurth},
  D.~{Jewitt}, \& K.~{Keil}, 231, \dodoi{10.48550/arXiv.astro-ph/0605597}

\bibitem[{{Ray} \& {Ferreira}(2021)}]{Ray2021}
{Ray}, T.~P., \& {Ferreira}, J. 2021, \nar, 93, 101615,
  \dodoi{10.1016/j.newar.2021.101615}

\bibitem[{{Reipurth} \& {Bally}(2001)}]{reipurthbally2001}
{Reipurth}, B., \& {Bally}, J. 2001, \araa, 39, 403,
  \dodoi{10.1146/annurev.astro.39.1.403}

\bibitem[{{Rigliaco} {et~al.}(2013){Rigliaco}, {Pascucci}, {Gorti}, {Edwards},
  \& {Hollenbach}}]{Rigliaco2013}
{Rigliaco}, E., {Pascucci}, I., {Gorti}, U., {Edwards}, S., \& {Hollenbach}, D.
  2013, \apj, 772, 60, \dodoi{10.1088/0004-637X/772/1/60}

\bibitem[{{Robinson} {et~al.}(2022){Robinson}, {Espaillat}, \&
  {Rodriguez}}]{Robinson2022}
{Robinson}, C.~E., {Espaillat}, C.~C., \& {Rodriguez}, J.~E. 2022, \apj, 935,
  54, \dodoi{10.3847/1538-4357/ac7e51}

\bibitem[{{Savage} \& {Mathis}(1979)}]{Savage1979}
{Savage}, B.~D., \& {Mathis}, J.~S. 1979, \araa, 17, 73,
  \dodoi{10.1146/annurev.aa.17.090179.000445}

\bibitem[{{Simon} {et~al.}(2016){Simon}, {Pascucci}, {Edwards}, {Feng},
  {Gorti}, {Hollenbach}, {Rigliaco}, \& {Keane}}]{Simon2016}
{Simon}, M.~N., {Pascucci}, I., {Edwards}, S., {et~al.} 2016, \apj, 831, 169,
  \dodoi{10.3847/0004-637X/831/2/169}

\bibitem[{{Takami} {et~al.}(2001){Takami}, {Bailey}, {Gledhill},
  {Chrysostomou}, \& {Hough}}]{Takami2001}
{Takami}, M., {Bailey}, J., {Gledhill}, T.~M., {Chrysostomou}, A., \& {Hough},
  J.~H. 2001, \mnras, 323, 177, \dodoi{10.1046/j.1365-8711.2001.04172.x}

\bibitem[{{Weber} {et~al.}(2020){Weber}, {Ercolano}, {Picogna}, {Hartmann}, \&
  {Rodenkirch}}]{Weber2020}
{Weber}, M.~L., {Ercolano}, B., {Picogna}, G., {Hartmann}, L., \& {Rodenkirch},
  P.~J. 2020, \mnras, 496, 223, \dodoi{10.1093/mnras/staa1549}

  \bibitem[{{Whelan} {et~al.}(2004){Whelan}, {Ray}, \& {Davis}}]{Whelan2004}
{Whelan}, E.~T., {Ray}, T.~P., \& {Davis}, C.~J. 2004, \aap, 417, 247,
  \dodoi{10.1051/0004-6361:20034381}

  \bibitem[{{Whelan} {et~al.}(2005){Whelan}, {Ray}, {Bacciotti}, {Natta},
  {Testi}, \& {Randich}}]{Whelan2005}
{Whelan}, E.~T., {Ray}, T.~P., {Bacciotti}, F., {et~al.} 2005, \nat, 435, 652,
  \dodoi{10.1038/nature03598}

\bibitem[{{Whelan} \& {Garcia}(2008)}]{Whelan2008}
{Whelan}, E., \& {Garcia}, P. 2008, in Jets from Young Stars II, ed.
  F.~{Bacciotti}, L.~{Testi}, \& E.~{Whelan}, Vol. 742 (Springer-Verlag:
  Berlin, Heidelberg), 123

\bibitem[{{Whelan} {et~al.}(2010){Whelan}, {Dougados}, {Perrin}, {Bonnefoy},
  {Bains}, {Redman}, {Ray}, {Bouy}, {Benisty}, {Bouvier}, {Chauvin}, {Garcia},
  {Grankvin}, \& {Malbet}}]{Whelan2010}
{Whelan}, E.~T., {Dougados}, C., {Perrin}, M.~D., {et~al.} 2010, \apjl, 720,
  L119, \dodoi{10.1088/2041-8205/720/1/L119}

\bibitem[{{Whelan} {et~al.}(2014{\natexlab{a}}){Whelan}, {Alcal{\'a}},
  {Bacciotti}, {Nisini}, {Bonito}, {Antoniucci}, {Stelzer}, {Biazzo}, {D'Elia},
  \& {Ray}}]{Whelan2014AA}
{Whelan}, E.~T., {Alcal{\'a}}, J.~M., {Bacciotti}, F., {et~al.}
  2014{\natexlab{a}}, \aap, 570, A59, \dodoi{10.1051/0004-6361/201424067}

\bibitem[{{Whelan} {et~al.}(2014{\natexlab{b}}){Whelan}, {Alcal{\'a}},
  {Bacciotti}, {Nisini}, {Bonito}, {Antoniucci}, {Stelzer}, {Biazzo}, {D'Elia},
  \& {Ray}}]{Whelan2014b}
{Whelan}, E.~T., {Alcal{\'a}}, J.~M., {Bacciotti}, F., et al.\ 2014{\natexlab{b}}, \aap, 570, A59, \dodoi{10.1051/0004-6361/201424067}


\bibitem[{{Whelan}(2014)}]{Whelan2014}
{Whelan}, E.~T. 2014, Astronomische Nachrichten, 335, 537,
  \dodoi{10.1002/asna.201412062}

\bibitem[{{Whelan} {et~al.}(2021){Whelan}, {Pascucci}, {Gorti}, {Edwards},
  {Alexander}, {Sterzik}, \& {Melo}}]{Whelan2021}
{Whelan}, E.~T., {Pascucci}, I., {Gorti}, U., {et~al.} 2021, \apj, 913, 43,
  \dodoi{10.3847/1538-4357/abf55e}


\bibitem[{{Whelan}(2023)}]{Whelan2023a}
{Whelan}, E.~T.\ 2023, Nature Astronomy, 7, 886, \dodoi{10.1038/s41550-023-01980-4}

\bibitem[{{Whelan} {et~al.}(2023){Whelan}, {Murphy}, \&
  {Pascucci}}]{Whelan2023b}
{Whelan}, E.~T., {Murphy}, A., \& {Pascucci}, I. 2023, \apj, 951, 1,
  \dodoi{10.3847/1538-4357/acd542}







\end{thebibliography}
\end{document}